\PassOptionsToPackage{pdfpagelabels=false}{hyperref}
\documentclass[fleqn,usenatbib]{mnras}

\usepackage[T1]{fontenc}
\usepackage{ae,aecompl}

\usepackage{graphicx}	
\usepackage{amsmath}	
\usepackage{amssymb}	
\usepackage{amsbsy}	    
\usepackage{subfig}
\usepackage{caption}

\def\tvi(#1,#2){\vrule height #1pt depth #2pt width 0pt}
\def\p{\partial}
\def\e{{\rm e}}
\def\d{{\rm d}}

\def\rso{r_{\rm s}}
\def\rco{R_{\rm c}}
\def\cs{c_{\rm s}}

\def\M{{\cal M}}

\title[Rotating collapse]{Impact of rotation on the evolution of convective vortices in collapsing stars}

\author[E. Abdikamalov, T. Foglizzo, O. Mukazhanov]{
E. Abdikamalov$^{1,2}$\thanks{E-mail: ernazar.abdikamalov@nu.edu.kz}
T. Foglizzo$^{3}$\thanks{E-mail: foglizzo@cea.fr}
and O. Mukazhanov$^{1}$\thanks{E-mail: olzhas.mukazhanov@nu.edu.kz}
\\
$^{1}$Department of Physics, Nazarbayev University, Nur-Sultan 010000, Kazakhstan\\
$^{2}$Energetic Cosmos Laboratory, Nazarbayev University, Nur-Sultan 010000, Kazakhstan\\
$^{3}$UMR AIM, CEA-CNRS-Univ. Paris Diderot, CEA Saclay, F-91191 Gif-sur-Yvette, France
}

\date{Accepted XXX. Received YYY; in original form ZZZ}

\pubyear{2020}

\begin{document}
\label{firstpage}
\pagerange{\pageref{firstpage}--\pageref{lastpage}}
\maketitle

\begin{abstract}
We study the impact of rotation on the hydrodynamic evolution of convective vortices during stellar collapse. Using linear hydrodynamics equations, we study the evolution of the vortices from their initial radii in convective shells down to smaller radii where they are expected to encounter the supernova shock. We find that the evolution of vortices is mainly governed by two effects: the acceleration of infall and the accompanying speed up of rotation. The former effect leads to the radial stretching of vortices, which limits the vortex velocities. The latter effect leads to the angular deformation of vortices in the direction of rotation, amplifying their non-radial velocity. We show that the radial velocities of the vortices are not significantly affected by rotation. We study acoustic wave emission and find that it is not sensitive to rotation. Finally, we analyze the impact of the corotation point and find that it has a small impact on the overall acoustic wave emission.  
\end{abstract}

\begin{keywords}
accretion, accretion discs -- convection -- hydrodynamics -- stars: massive -- supernovae: general -- turbulence
\end{keywords}


\section{Introduction}

Convection in the innermost shells of massive stars may have a profound impact on the core-collapse supernova (CCSNe) explosions that these stars may produce \citep{couch:13d, couch:15b, mueller:16, mueller:17, takahashi:16, nagakura:19}. We can look at this mechanism as a sequence of several processes. First, upon reaching its maximum mass, the iron core becomes unstable and starts collapsing. The convective shell follows the iron core and starts descending too \citep{takahashi:14, lai:00, buras:06b}. However, the shells initially descend at a slower pace than the core due to smaller free-fall velocity at larger radii. The core bounces ${\sim} 300 \,\mathrm{ms}$ later, launching a shock wave. This shock stalls at a radius of ${\sim} 150\,\mathrm{km}$ within a few tens of milliseconds after formation. The convective shell encounters the shock ${\sim} 200\,\mathrm{ms}$ later \citep{mueller:17}. The interaction of the convective vortices with the shock generates perturbations in the post-shock flow \citep{abdikamalov:16, abdikamalov18, huete:18, huete:19}, which amplify the turbulence in that region, pushing the shock forward. Large-scale perturbations were found to have particularly strong impact on the shock expansion \citep{mueller:16, kazeroni20}. Such perturbations are likely to originate in the oxygen burning and, to a lesser extent, silicon burning shells of massive stars \citep{collins:18}. See \citet{mueller:20review} for a comprehensive recent review. 

In our previous work \citep{abdikamalov20}, we studied the evolution of convective vortices during stellar collapse for non-rotating stars. Our main aim was to establish the physical nature and the qualitative parameters of the vortices when they reach the inner regions of the flow, where they are expected to encounter the supernova shock. We found that the acceleration of the infall plays a crucial role in the evolution of the vortices. The acceleration stretches the vortices in the radial direction. The stretching of the vortices leads to a decrease of their velocities. In addition, the vortex motion distorts isodensity surface, leading to pressure perturbations and the generation of acoustic waves \citep{kovalenko:98, mueller:15}. Upon reaching the supernova shock, the velocity perturbations associated to acoustic waves are likely to be much stronger than those associated to vortices. In this work, we extend these results to incorporate the rotation of the stellar core. 

As in \cite{abdikamalov20}, we use an idealized model based on the solution of linear hydrodynamics equation on a stationary background flow. This approach has a particular advantage: due to our targeted approximations, we can disentangle the impact of rotation from other processes that are less important to the dynamics. Note that we do not study how rotation influences convection during stellar evolution prior to core collapse. Instead, we explore how rotation affects the hydrodynamic evolution of vortices during core-collapse. We find that, similarly to the non-rotating case, the radial stretching due to the accelerated infall plays a crucial role in the evolution of the vortices. In addition, we find that the speed-up of rotation during the infall deforms the vortices in the direction of rotation. As a result, the vortices acquire stronger non-radial velocities. However, the radial component is not sensitive to rotation. Since acoustic waves are generated mostly by radial distortions, acoustic wave emission is not sensitive to rotation. We investigate the role of the corotation point, where the pattern speed of the vortices matches the rotation speed, and find that it has little impact on the overall acoustic emission. 

The ultimate motivation behind our work is to contribute to better understanding the explosion mechanism of rotating stars. While the explosion of non-rotating stars have been studied extensively \cite[e.g.,][for recent reviews]{janka:12a, mueller:20review}, relatively little is understood about the explosions of rotating stars. The fastest rotating stars produce millisecond-period neutron stars and power supernovae (aka hypernovae) with ${\sim}10^{52}\,\mathrm{erg}$ energies \citep{meier:76, bisno:76, burrows:07b}. In these rare stars \citep{heger:05, popov:12, mosser:12, deheuvels:14, cantiello:14}, the magnetic fields are expected to dominate the explosion dynamics by transferring the rotation kinetic energy of the proto-neutron star to the explosion front \citep{akiyama:03, winteler12, moesta:14b, kuroda20, obergaulinger20a, raynaud20}. In slower rotating stars, a combination of neutrino heating, neutrino-driven  multi-dimensional hydrodynamic instabilities, as well as magnetic fields may help to produce explosions \citep{endeve:10, endeve:12, nakamura:14, summa:18, fujisawa19, mueller20b}. This is the regime where the pre-collapse convection could be important to the explosion dynamics. 

This paper is organized as follows. In Section~\ref{sec:method} we describe our methods. In Section~\ref{sec:results} we present our results. Finally, in Section~\ref{sec:conclusion} we provide our conclusions. 

\section{Method}
\label{sec:method}

\begin{figure*}
    \centering
    \includegraphics[width=0.95\textwidth]{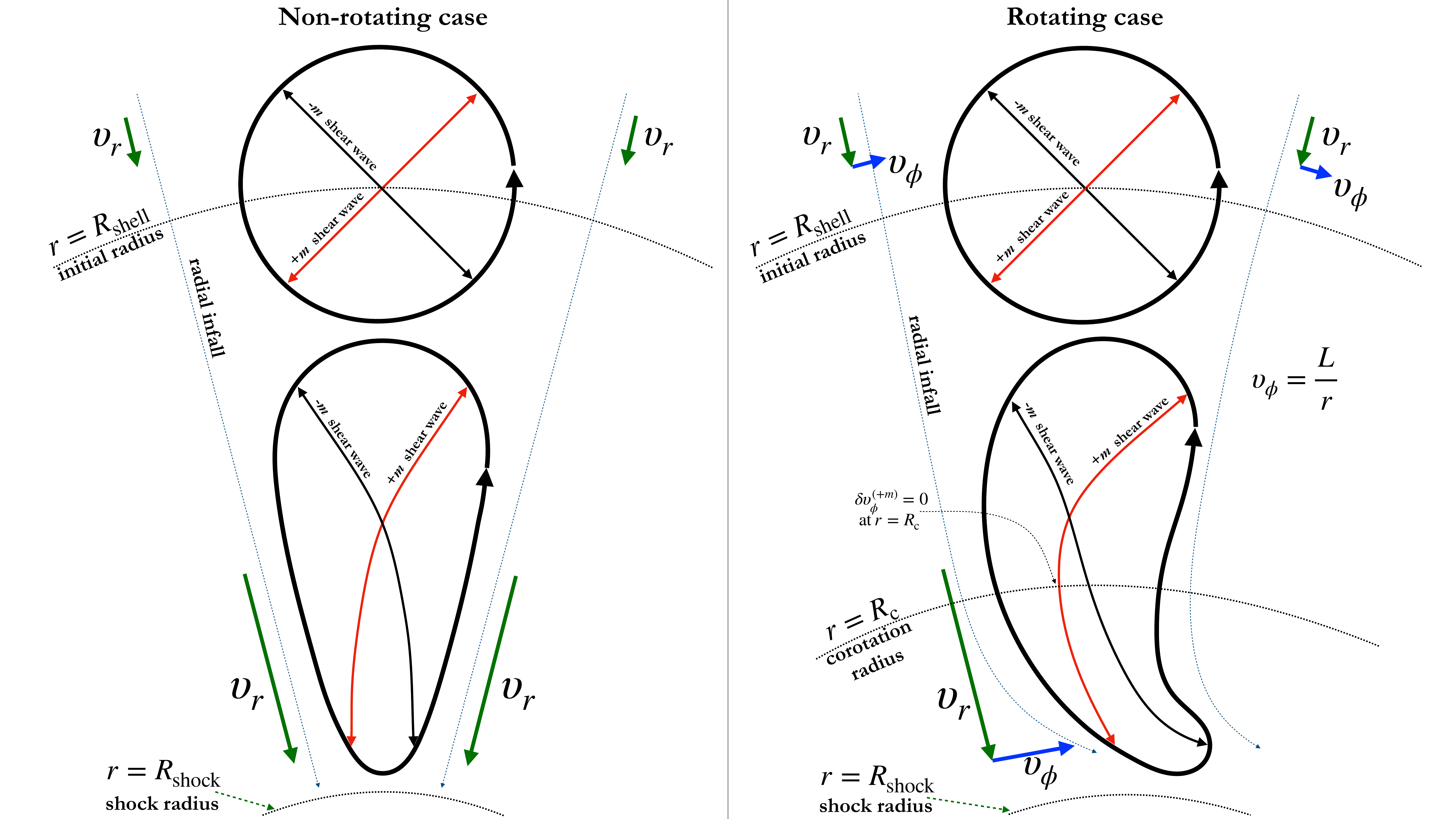}
    \caption{Schematic depiction of the evolution of convective vortices in non-rotating (left panel) and rotating (right panel) cases. $R_\mathrm{shell}$ is the initial radius of the convective shell, while $R_\mathrm{shock}$ is the radius where the convective vortices are expected to encounter the supernova shock. A vortex with wavenumber $m$ can be represented as a superposition of two shear waves with wavenumbers $m$ and $-m$. Due to the accelerated collapse, the vortex undergoes a radial elongation. The differential rotation deforms it in the positive $\phi$ direction.} 
    \label{fig:scheme}
\end{figure*}

We treat convective vortices as linear perturbations resulting from perturbations of velocity in the equatorial plane of an axisymmetric background flow that represents a collapsing rotating star. For brevity, we hereafter refer to the vorticity perturbations as vortices. The flow is assumed to be adiabatic during the collapse, which is a reasonable approximation as neutrino cooling has a longer timescale than collapse \cite[e.g.,][]{yamamoto:13}. We also neglect nuclear burning as it is unlikely to affect the convective dynamics during collapse due to the longer convective turnover timescale \cite[e.g.,][]{mueller:16}. We model the stellar matter with an ideal gas equation of state with {adiabatic index} $\gamma=4/3$, which is a good approximation for the radiation-dominated stellar gas \citep[e.g.,][]{arnett:96}. We also assume that the flow is isentropic. However, advection of entropy perturbations can generate both vorticity and acoustic waves \citep{abdikamalov20}. This will be studied for rotating collapse in our subsequent work. 

The stationary background flow is given by the transonic Bondi solution with rotation. This solution is characterized by a sonic radius $\rso$, above (below) which the collapse velocity is subsonic (supersonic). Since angular momentum is conserved during stellar core-collapse, we assume constant specific angular momentum. For a given $\gamma$, the background solution depends on the mass of the central gravitating core, rotation, and fluid pressure. We assume that the mass of the central source is $1.4M_\odot$, which is in line with the predictions from 3D simulations \citep[e.g.,][]{nagakura20}. Because the sonic radius $\rso$ depends on the sound speed and thus on temperature, we can fix the scale of the fluid pressure by selecting the value of $\rso$. We adopt $\rso=1{,}500\,\mathrm{km}$, which is a ballpark figure for the early postbounce phase \citep[e.g.,][]{takahashi:14}. 

Our model incorporates the first order effects of rotation on the structure of the collapsing star. This includes the slow-down of collapse due to the centrifugal force and the differential rotation resulting from the conservation of angular momentum. 
We solve the equations on the equatorial plane using spherical coordinates, as described in Appendix~\ref{sec:formalism}. 
The structure of the flow in the direction perpendicular to the equatorial plane is simplified by neglecting the poloidal derivatives in both the background flow and the perturbations. 
The centrifugal deformation of the star to an oblate shape, which is a higher-order effect \citep[e.g.,][]{tassoul:78}, is not included. This is justified because, as we will see below, the ratio of the rotational kinetic energy to potential binding energy, which measures the relative importance of rotation, remains below ${\sim} 16\%$ in the region of interest even for the most extreme rotation that we consider.
The fastest rotating stars are expected to have pre-supernova angular momenta of ${\lesssim} 10^{16}\,\mathrm{cm^2/s}$ in their cores \citep{woosley:06}. In our work, we adopt a somewhat higher value of $L_\mathrm{max}=3\times 10^{16}\,\mathrm{cm^2/s}$ as a crude upper limit for the angular momentum. Note that only a tiny minority (perhaps ${\sim}1\%$) of stars are expected to have angular momenta approaching this limit in their cores in the pre-supernova stage \cite[e.g.,][]{woosley:06}. We consider $10$ different values of $L$ ranging from $0$ to $L_\mathrm{max}$ with a step of $0.1 L_\mathrm{max}$. 

The solution procedure for obtaining the background solution is explained in Appendix~\ref{sec:stationary_solution}. 

The evolution of the vortices and the acoustic waves is governed by linear hydrodynamics equations, which we formulate in a compact form as a second-order inhomogeneous ordinary differential equation (cf. Appendix~\ref{sec:perturbation_solution} for its derivation),
\begin{flalign}
\left\lbrace{\p^2 \over \p X^2}+W\right\rbrace( r\delta \tilde \upsilon_\phi)=-
\e^{\int {i\omega'\over c^2}\d X}{\p \over \p X}{r\delta w_\theta\over \upsilon_r}, \label{rdvtilde1} 
\end{flalign}
where $X$ is related to the radial coordinate $r$ (via Eq.~\ref{eq:X}), while $\delta \tilde \upsilon_\phi$ is related to the $\phi$ velocity perturbation (via Eq.~\ref{eq:vphitilde}). The dependence on angle $\phi$ is separated by assuming $\exp(im\phi)$ dependence, where $m$ is the angular wavenumber of the perturbation. The time-dependence is separated by assuming $\exp(-i\omega t)$, where $\omega$ is the frequency. Here, $\delta w_\theta$ is the $\theta$-component of the vorticity, i.e., the vortex motion along the equatorial plane. The Doppler shifted frequency $\omega'$ is defined as \citep{yamasaki:08}
\begin{equation}
\label{eq:omegap}
\omega' = \omega - \frac{m L}{r^2}
\end{equation}
Eq.~(\ref{rdvtilde1}) is remarkably similar to its counterpart in the non-rotating case. The impact of rotation is contained in the variable $\omega'$. In the non-rotating limit, $\omega'$ reduces to $\omega$. Hence, we can apply the methodology that we developed for the non-rotating case \citep{abdikamalov20} with just a minor generalization to the non-uniform Doppler-shifted frequency. As in the non-rotating case, the boundary conditions are imposed by assuming no incoming acoustic waves from infinity and regularity at the sonic point. We obtain a homogeneous solution that is regular at the sonic point using the Frobenius expansion \citep{foglizzo:01}. The in-going and out-going acoustic waves are identified at the outer boundary of the computational domain using the Wentzel-Kramers-Brillouin (WKB) method (cf. Appendix~\ref{sec:refraction_coeff}). The outer boundary is chosen at $40\rso$, which is sufficient for the validity of the WKB approximation, as demonstrated in \citet{abdikamalov20}. 

A circular motion can be represented as a superposition of two harmonic oscillations along two orthogonal directions. Similarly, a vortex comoving with the fluid can be modeled as two shear waves that are inclined by $45^\circ$ and $-45^\circ$ with respect to the mean flow direction. Using this normalization, we model a vortex with wavenumber $m$ as a combination of two shear waves with wavenumbers $m$ and $-m$. This is schematically depicted in the upper left part of Fig.~\ref{fig:scheme}. 

We parametrize the vortices by their angular wavenumber $m$ and their initial radius $R_\mathrm{shell}$. We consider 4 different values of $R_\mathrm{shell}$ ranging from $\rso$ to $4\rso$, which is comparable to the radii of the oxygen and silicon shells \cite[e.g.,][]{mueller:16}. We evolve perturbations down to the radius $0.1\rso=150\,\mathrm{km}$, i.e. where the stalled shock is expected to reside when reached by the innermost convective shell. The vortices are assumed to be close to circular at their initial radii. This condition is expressed as the equality of the $r$ component of the velocity (Eq.~\ref{eq:vr_vor}) of the shear wave with wavenumber $m$ to the $\phi$ velocity component (Eq.~\ref{eq:vphi_vor}) of the $-m$ shear wave. We normalize vortex velocities to obtain a Mach number of $0.1$ at the initial radius $R_\mathrm{shell}$, in agreement with numerical simulations \citep{collins:18, yoshida19, yadav20, chatzopoulos:14}. Due to the linearity of our formalism, our results can be scaled linearly to any other value of the initial Mach number.

As discussed above, in our model we make a number of targeted approximations: we assume spherical stationary adiabatic flow with constant entropy and neglect neutrino cooling as well as nuclear reactions. These limitations preclude quantitatively precise results but that is not our aim. Our goal is to investigate the effect of rotation on the hydrodynamic evolution of vortices during their infall towards the supernova shock.

\section{Results}
\label{sec:results}

\subsection{Stationary background solution}
\label{sec:background}

\begin{figure}
    \centering
    \includegraphics[width=0.5\textwidth]{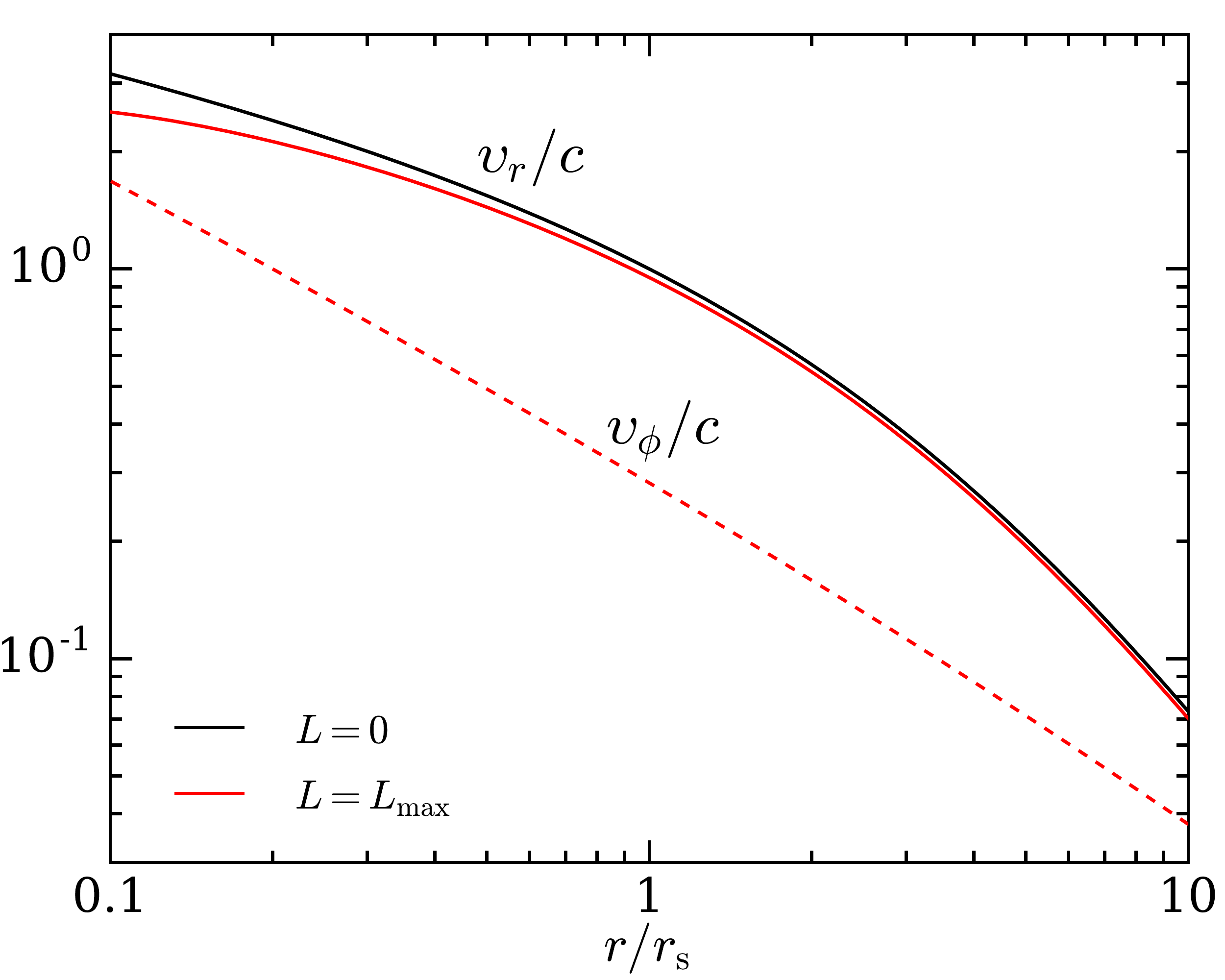}
    \caption{Radial infall speed $v_r$ (solid lines) and rotation speed $v_\phi$ (dashed lines) in units of the local sound speed $c$ for rotating Bondi solution as functions of $r/\rso$ for non-rotating (black lines) and extremely rapidly rotating (red lines) cases. Here, $\rso$ is the sonic radius and $L$ is the angular momentum of the flow. The maximum angular momentum $L_\mathrm{max}$ is taken to be $3{\times} 10^{16}\,\mathrm{cm^2/s}$. The azimutal velocity is given by $\upsilon_\phi=L/r$. The stellar matter is modeled as $\gamma=4/3$ ideal gas EOS.}
    \label{fig:back}
\end{figure}

We start our analysis by looking at the impact of the stellar rotation on the stationary background solution. Fig.~\ref{fig:back} shows the radial profiles of the infall and rotation velocities of the non-rotating model (black line) and the extremely rapidly rotating model with $L=L_\mathrm{max}$ (red line). Fig.~\ref{fig:back} reveals that the rapidly rotating model does not differ much from the non-rotating model at large radii (e.g., at $r\gtrsim \rso$). This is expected since at large radii gravity is stronger than the centrifugal force. As the star collapses, the conservation of angular momentum forces the angular velocity $\Omega$ to increase as $1/r^2$. This means that the centrifugal force $\Omega^2r$ increases as $1/r^3$. Gravity, on the other hand, increases as $1/r^2$. Therefore, the centrifugal force becomes relatively important only at small radii. The relative importance of rotation at a given radius $r$ can be quantified using the ratio of the rotational kinetic energy $\mathrm{T}$ to gravitational binding energy $\mathrm{W}$ at that radius: 
\begin{eqnarray}
    \frac{\mathrm{T}}{|\mathrm{W}|}\sim 0.016
    \left(\frac{L}{L_\mathrm{max}}\right)^2 \left( \frac{\rso}{r} \right).
\end{eqnarray}
For example, at radius $r=0.2\rso$, this ratio becomes $\sim 0.06$ for the most rapidly rotating model with $L=L_\mathrm{max}$. In line with this scaling, the radial velocity of the rotating model is $7\%$ smaller than that of the non-rotating model at $0.2 \rso$. At $0.1 \rso$, the difference becomes as large as $18\%$. In the following, we will explore how these differences impact the evolution of the convective vortices during the stellar collapse.


\subsection{Vortex evolution}
\label{sec:vorticity}

\begin{figure*}
    \centering
    \includegraphics[width=0.45\textwidth]{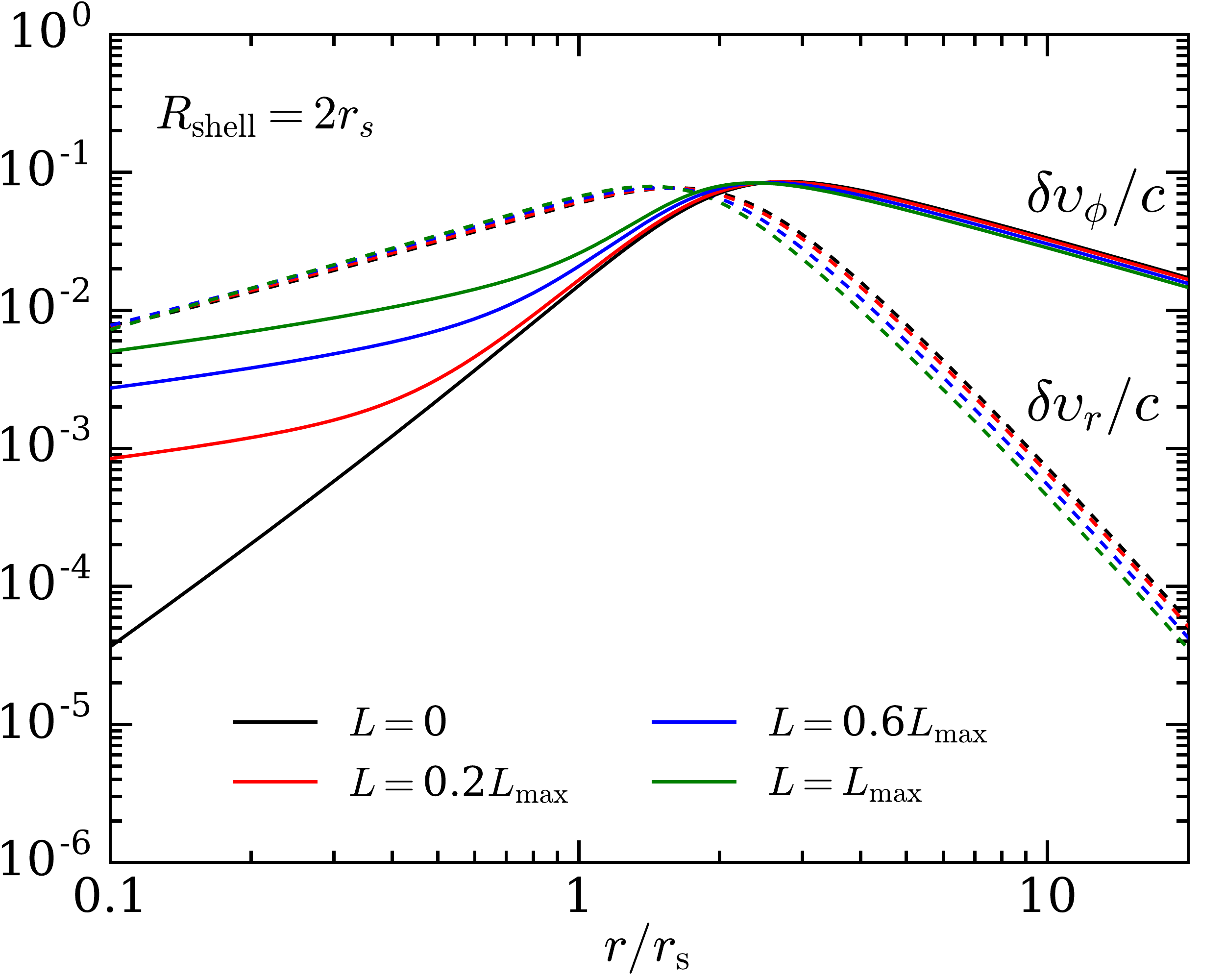}
    \includegraphics[width=0.45\textwidth]{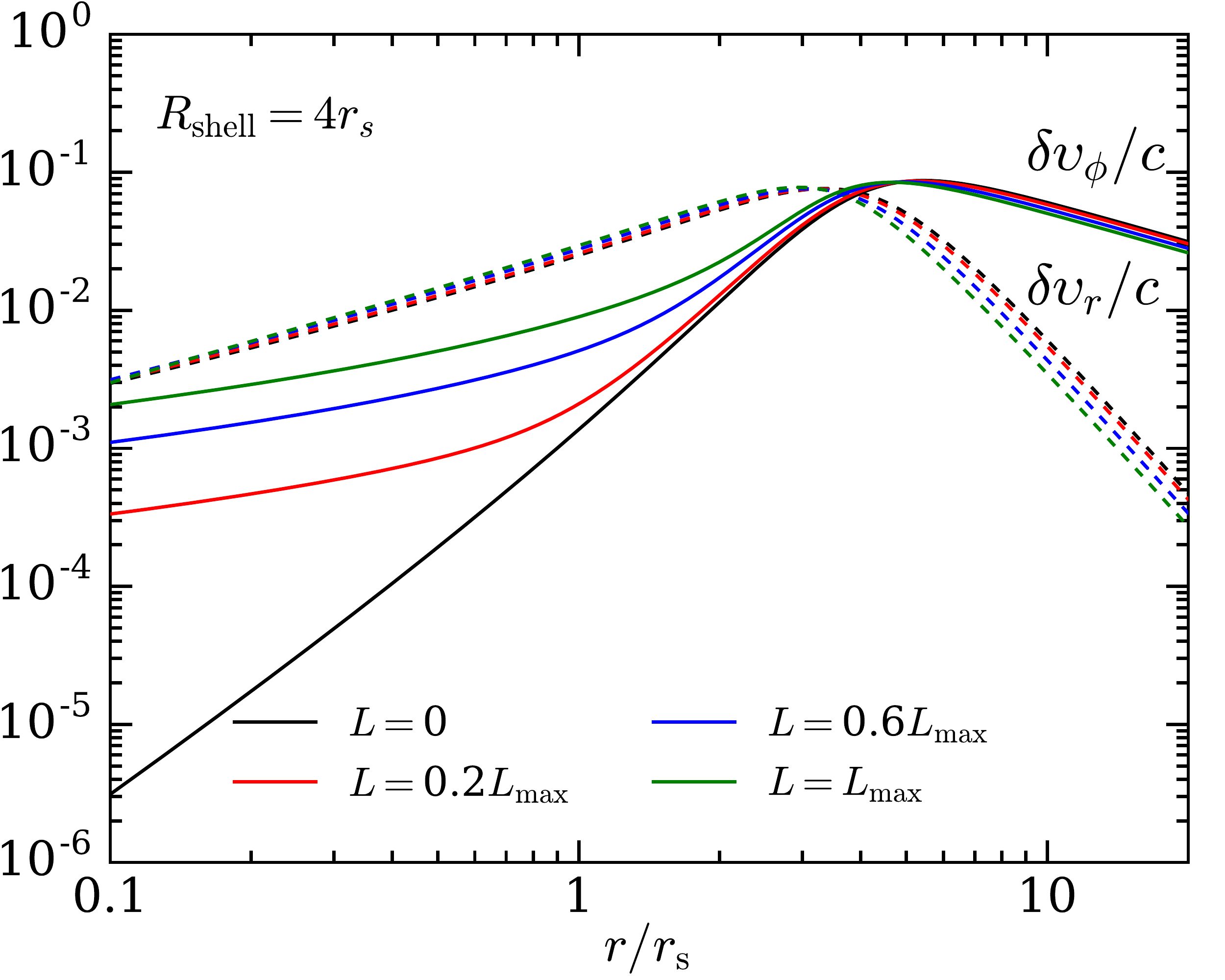}
    \caption{The radial and tangential velocities of the vortices as a function of radius for different values of angular momentum $L$. The left and right panels correspond to the vortices with initial radii of $R_{\rm shell}=2\rso$ and $R_{\rm shell}=4\rso$. The tangential component is larger for models with faster rotation, while the radial component is not sensitive to rotation. The perturbations are normalized to yield a convective Mach number of $0.1$ at $r=R_\mathrm{shell}$. With this normalization, the values of the velocities are independent of $m$. $L_\mathrm{max}=3{\times} 10^{16}\,\mathrm{cm^2/s}$ is the maximum angular momentum considered in this work, as discussed in Section~\ref{sec:method}.}
    \label{fig:vort_comb}
\end{figure*}

\begin{figure}
    \centering
    \includegraphics[width=0.45\textwidth]{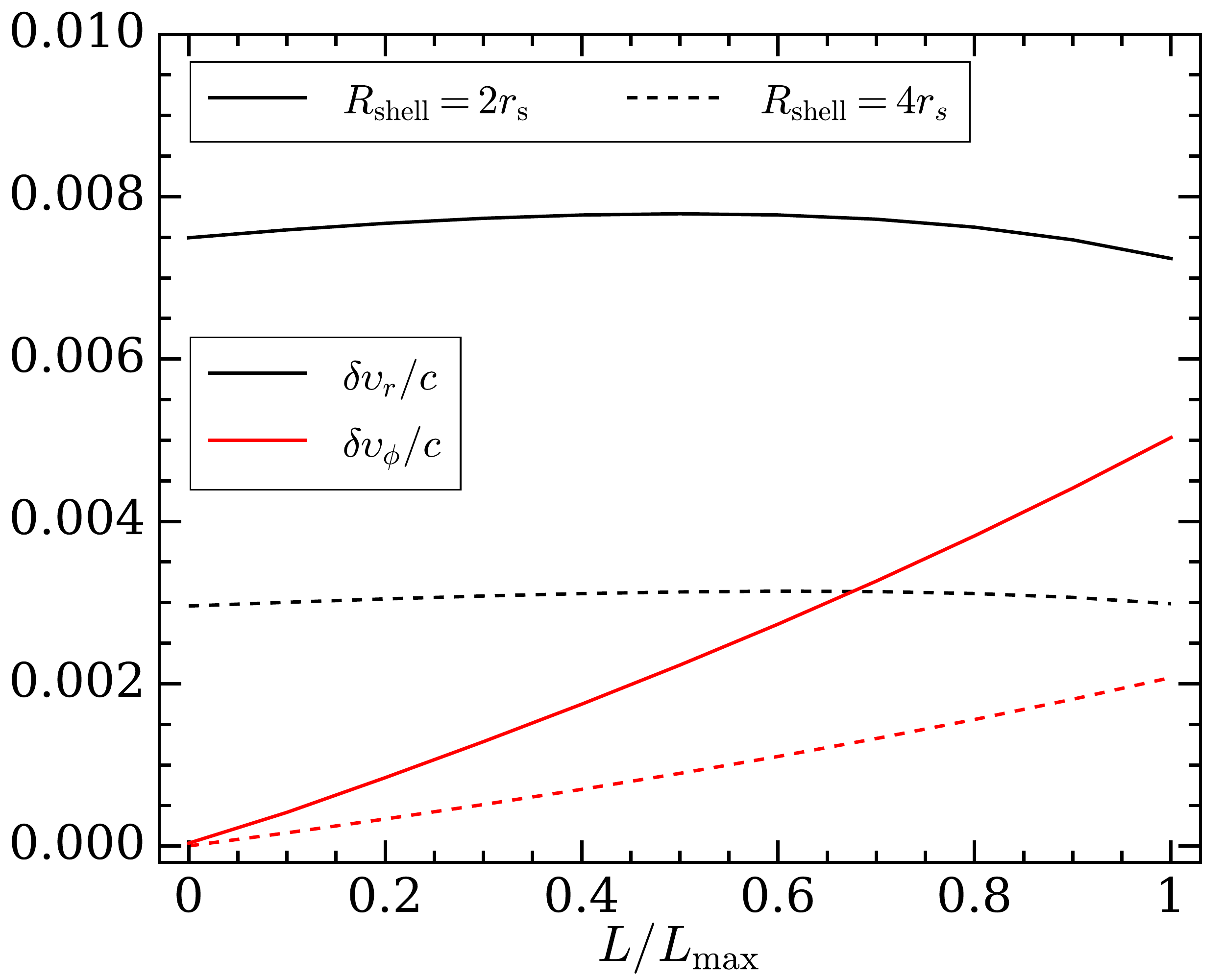}
    \caption{The radial and tangential velocity of the vortices at $r = 0.1r_s$ as a function of specific angular momentum $L$. The radial velocity is not sensitive to rotation, while the $\phi$ component increases with rotation. $L_\mathrm{max}=3{\times} 10^{16}\,\mathrm{cm^2/s}$ is the maximum angular momentum considered in this work, as discussed in Section~\ref{sec:method}.}
    \label{fig:vort_rmin}
\end{figure}

Figure~\ref{fig:vort_comb} shows the radial and tangential velocities $\delta \upsilon_r$ and $\delta \upsilon_\phi$ in units of the local sound speed $c$ for the vortices that originate at radii $R_\mathrm{shell}=2\rso$ (left panel) and $R_\mathrm{shell}=4\rso$ (right panel) as a function of radius for different values of $L$. As the vortices descend from their initial radius, both velocity components decrease with $r$. As established in \cite{abdikamalov20}, this is caused by the acceleration of the infall. The inner part of the vortices collapses faster than the outer part, leading to radial stretching of the vortices, as depicted schematically in Fig.~\ref{fig:scheme}. The circulation of the vortex lines, 
\begin{equation}
\Gamma=\oint {\delta \boldsymbol{\upsilon}} \cdot d {\bf s},
\end{equation}
is a conserved quantity for isentropic flows \citep[e.g.,][]{landau:59}. Therefore, the radial stretching of a vortex reduces its velocity. Due to this effect, the velocities of the vortices do not exceed $10^{-2}c$ by the time they reach the shock at $R_\mathrm{shock}=0.1\rso$. Since larger vortices have bigger differences in the accelerations in their innermost and outermost points, they experience stronger radial stretching. For example, the velocity of the vortex with $R_\mathrm{shell}=4\rso$ is ${\sim} 3$ times smaller than that with $R_\mathrm{shell}=2\rso$ at $0.1\rso$ as a result of the stronger stretching.  

As we can see in Fig.~\ref{fig:vort_comb}, the $\phi$ component $\delta \upsilon_\phi/c$ is larger in models with faster rotation, especially in the inner regions of the flow. We can see this trend also in Fig.~\ref{fig:vort_rmin}, which shows $\delta \upsilon_\phi/c$ as a function of $L$ at radius $0.1\rso$. For example, $\delta \upsilon_\phi/c$ at $0.1\rso$ for $L=L_\mathrm{max}$ is ${\sim} 10^2$ times larger than that for $L=0$. The cause of this behavior can be illustrated with the help of the schematic depiction on the right panel of Fig.~\ref{fig:scheme}. Since the rotation speed increases with decreasing $r$, the inner part of the vortex rotates faster than the outer part. The difference in the velocities becomes larger as the vortices become radially elongated due to the accelerated infall. The faster rotation at the inner point deforms the vortex in the $\phi$ direction. As a result, the vortex velocity vector acquires stronger non-radial component.

On the other hand, the radial velocity of the vortices is remarkably insensitive to rotation. For example, at $r=0.1\rso$, the variations in $\delta \upsilon_r/c$ remain below ${\sim}5\%$ when $L$ changes from $0$ to $L_\mathrm{max}$ (cf. Fig.~\ref{fig:vort_rmin}). Rotating models collapse slower, leading to smaller radial stretching. At the same time, the deformation of the vortices in the angular direction is stronger in such models. These two effects have opposite impact on $\delta \upsilon_r/c$, contributing to the weak dependence of the latter on rotation. 


\begin{figure*}
    \centering
    \includegraphics[width=0.45\textwidth]{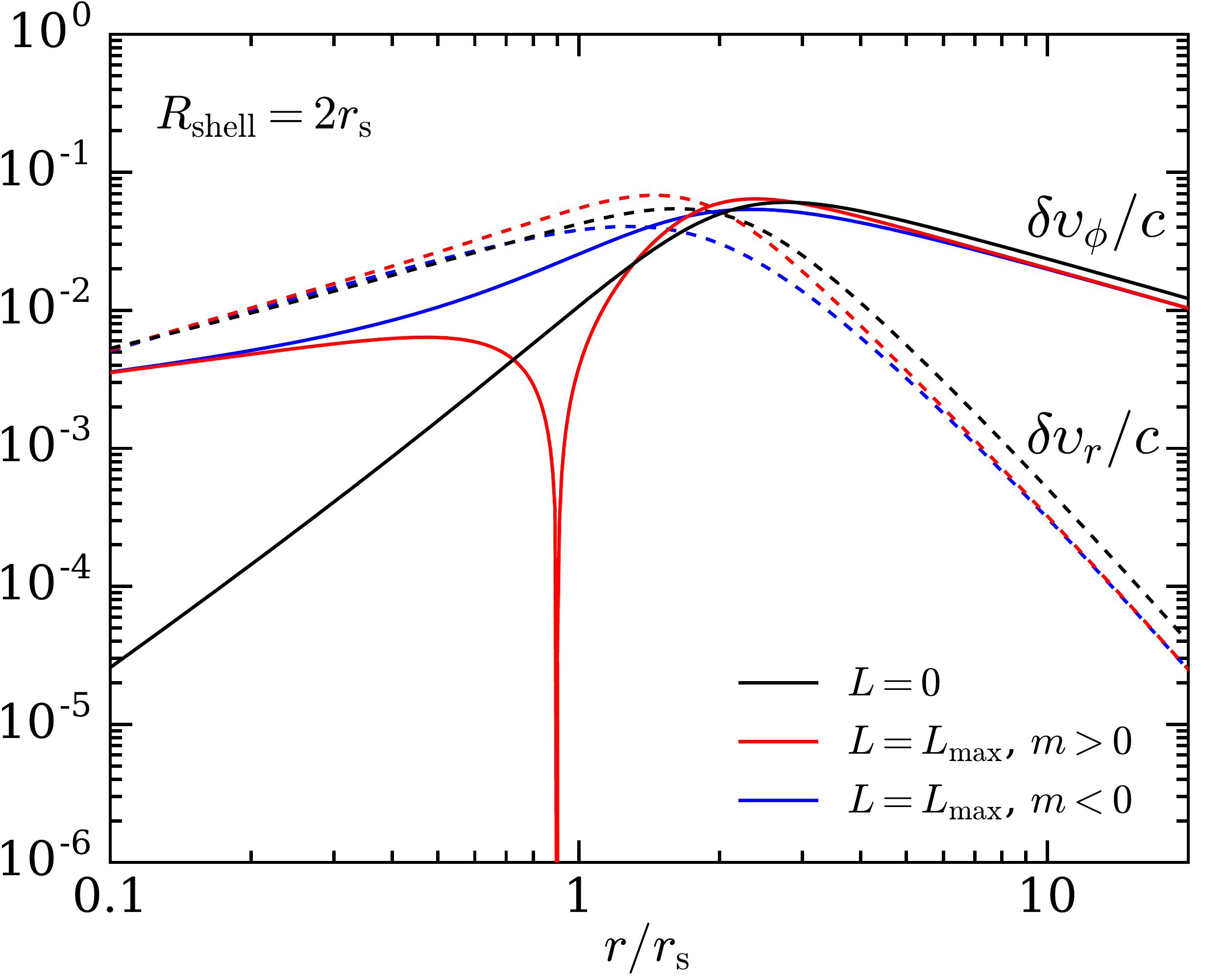}
    \includegraphics[width=0.45\textwidth]{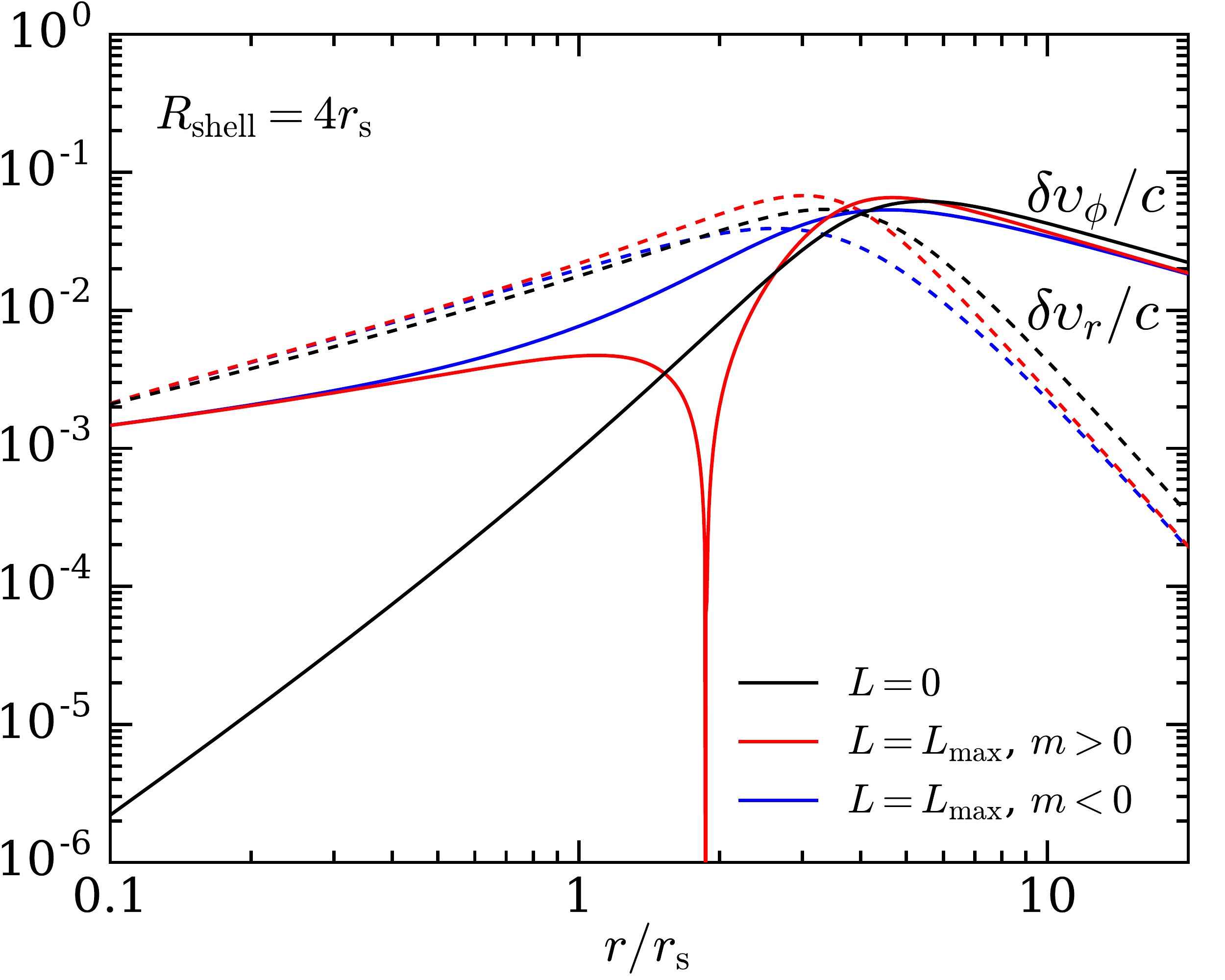}
    \caption{The radial and tangential velocity of advected shear perturbations with positive and negative angular wavenumbers $m$ as a function of radius for different values of angular momentum $L$. The left and right panels correspond to shear waves with initial radii $R_{\rm shell}$ of $2\rso$ and $4\rso$. The tangential component increases with rotation, while the radial component is not affected. $L_\mathrm{max}=3\times 10^{16}\,\mathrm{cm^2/s}$ is the maximum angular momentum considered in this work, as discussed in Section~\ref{sec:method}.}
    \label{fig:vort_mm}
\end{figure*}

In order to further illustrate this effect and to acquire further insight into the behavior of the vortices, we look at the evolution of their constituent shear waves. As discussed above, a vortex with angular wavenumber $m$ can be expressed as a superposition of two shear waves wavenumbers $m$ and $-m$. Fig.~\ref{fig:vort_mm} shows $\delta \upsilon_\phi/c$ and $\delta \upsilon_r/c$ for these waves as a function of $r$ for non-rotating and rapidly rotating case with $L=L_\mathrm{max}$. The behavior of the shear waves mirrors the behavior of the vortices discussed previously. In the rotating case, both shear waves acquire strong non-radial velocity $\delta \upsilon_\phi/c$ at small radii. As illustrated in the schematic plot Fig.~\ref{fig:scheme}, the increase of the non-radial component is caused by the deformation of the radially-elongated shear waves in the direction of rotation.

Further analysis of Fig.~\ref{fig:vort_mm} reveals that the $\delta \upsilon_\phi/c$ of the shear wave with positive $m$ completely vanishes at a specific radius. This radius corresponds to the so-called corotation radius $R_\mathrm{c}$, a point where the pattern speed of the wave matches the rotation speed. At this point, the Doppler-shifted frequency~(\ref{eq:omegap}) becomes zero, which yields
\begin{eqnarray}
\label{eq:rc}
R_\mathrm{c} = \left(\frac{mL}{\omega}\right)^{1/2}
\end{eqnarray}
The behavior of velocity $\delta \upsilon_\phi/c$ at $R_\mathrm{c}$ can be explained using the schematic depiction in Fig.~\ref{fig:scheme}. The velocity vector of the shear wave is initially inclined by about $ 45^\circ$ with respect to the positive $r$ and $\phi$ axes. As the shear wave advects towards the center, it deforms in the $\phi$ direction due to the faster rotation in the inner regions. At the corotation radius, it becomes vertical and the $\phi$ velocity completely vanishes. The shear mode with negative $m$ does not exhibit such a behavior and it does not possess a corotation radius. Despite this difference at the corotation radius, the velocity perturbation of the two shear waves become similar again at small radii (${\sim} 0.1\rso$). 

\subsection{Generation of acoustic waves}
\label{sec:acoustic}

\begin{figure*}
    \centering
    \includegraphics[width=0.32\textwidth]{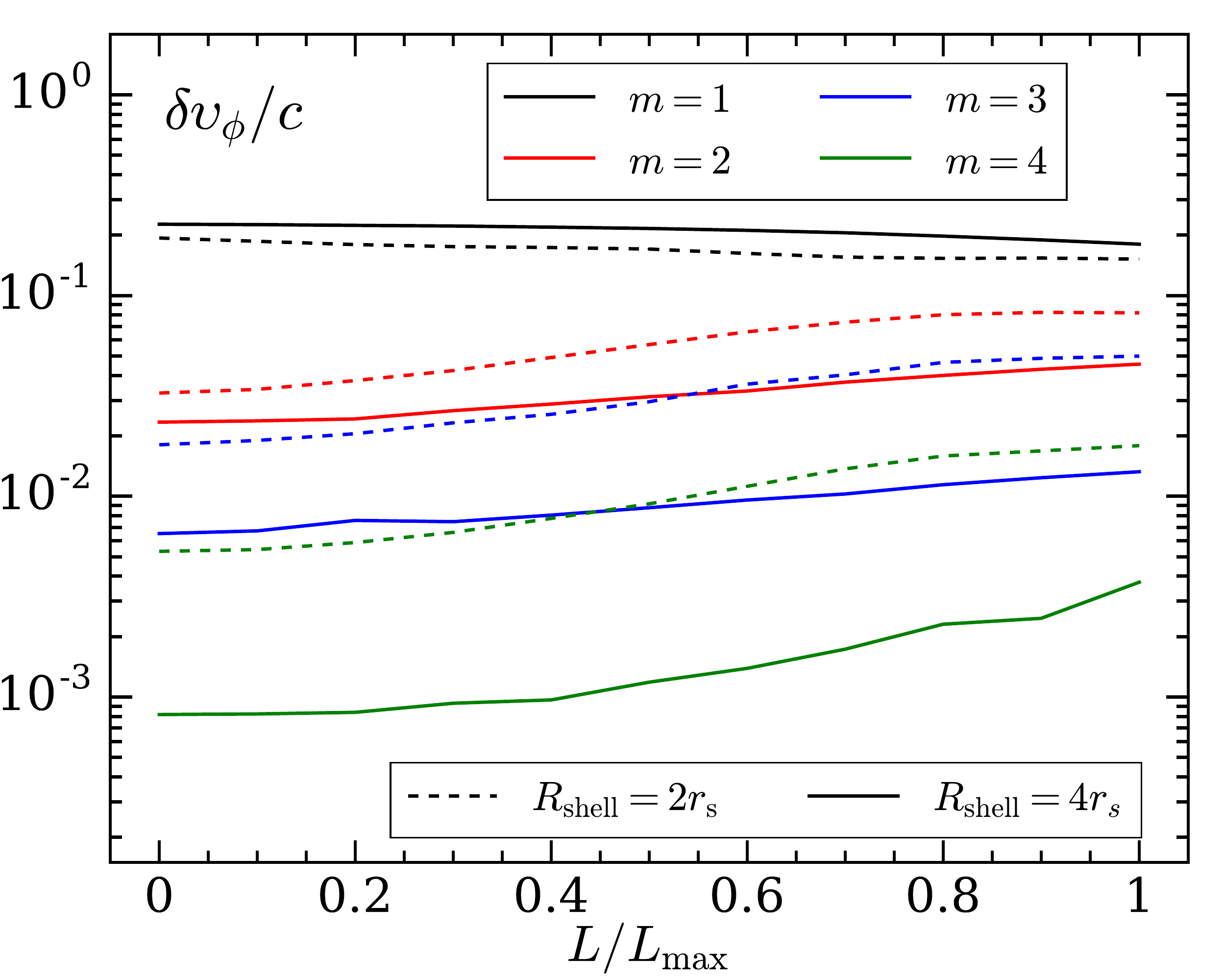}
    \includegraphics[width=0.32\textwidth]{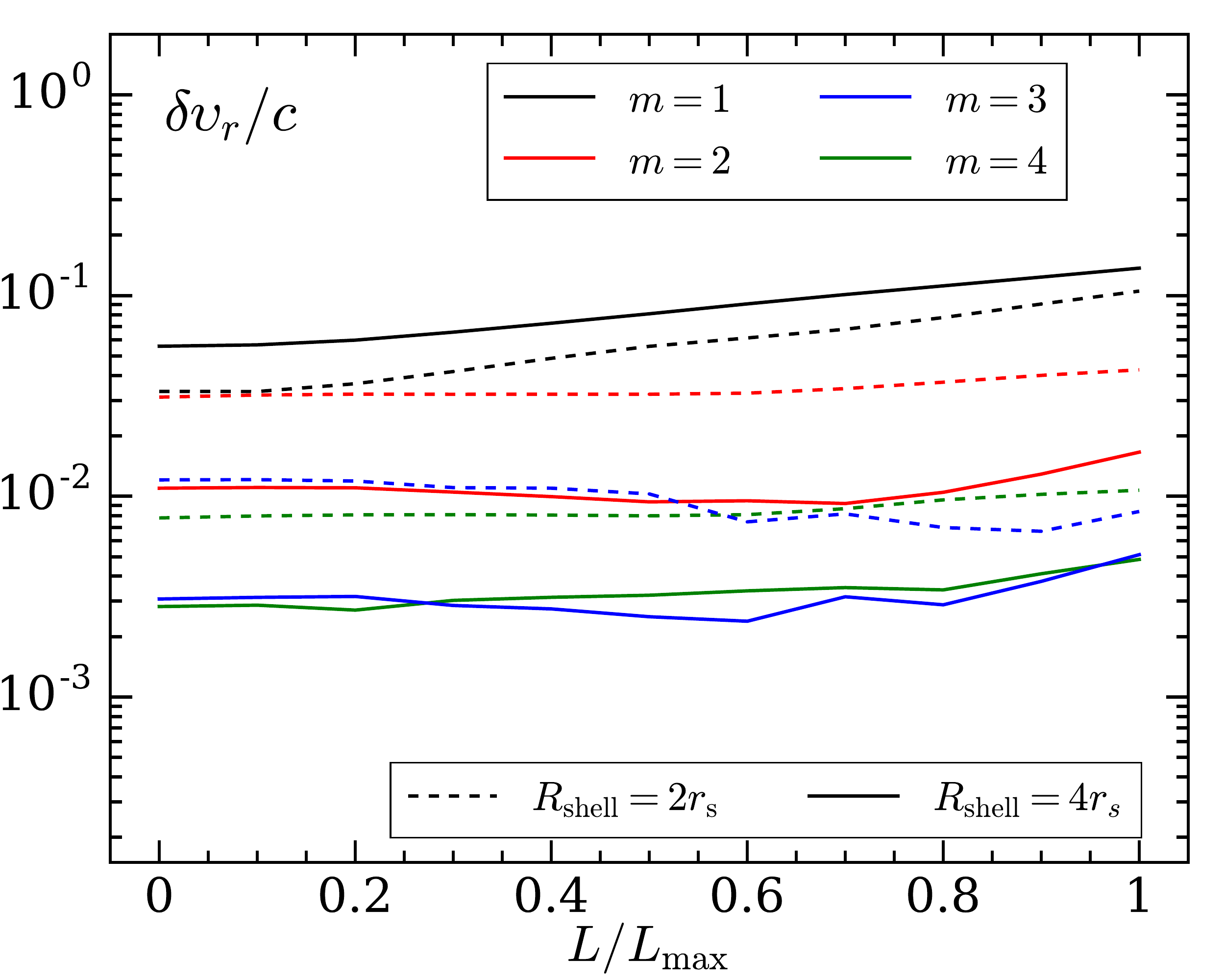}
    \includegraphics[width=0.32\textwidth]{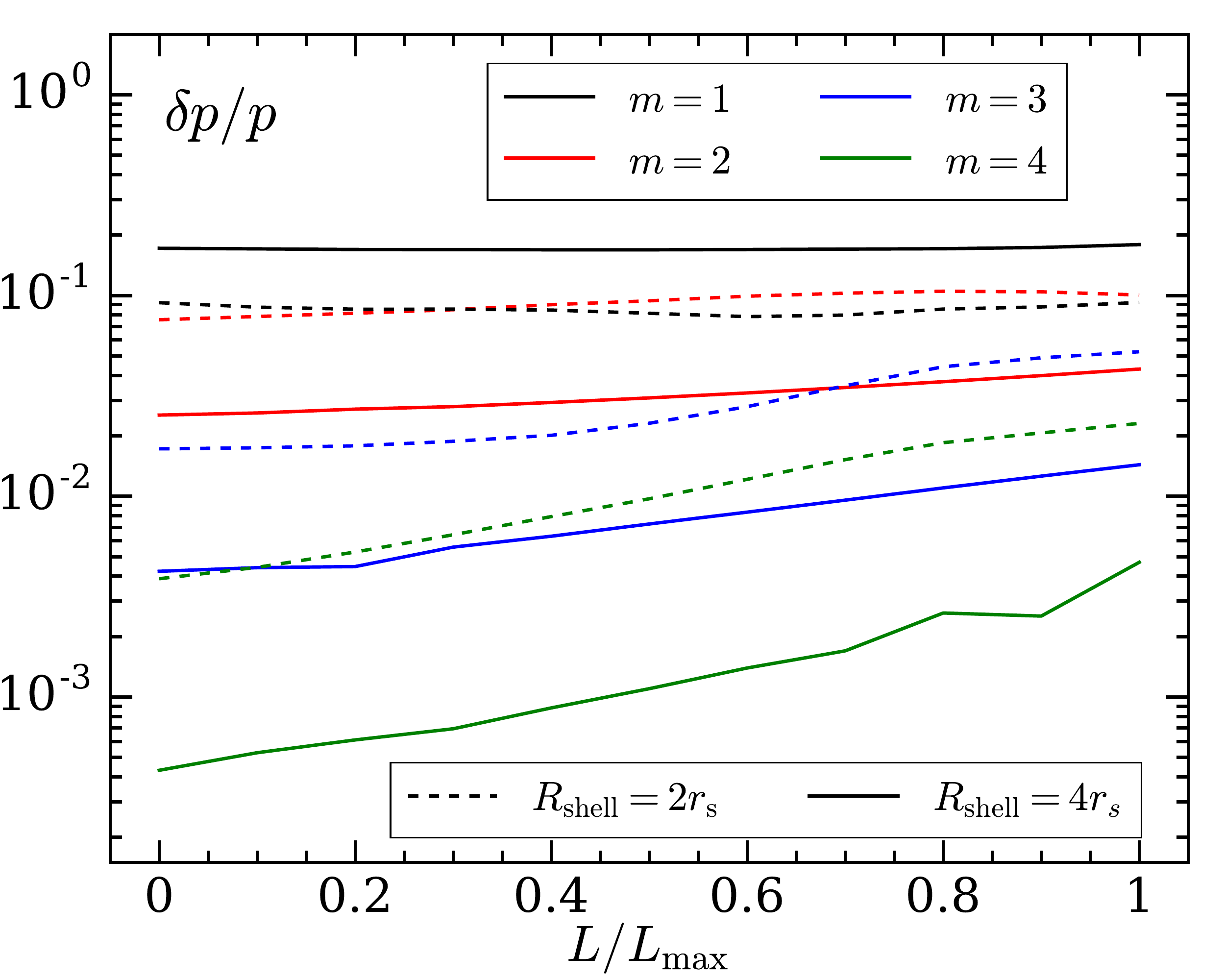}
    \caption{Velocity perturbations $\delta \upsilon_\phi/c$ (left panels) and $\delta \upsilon_r/c$ (center panels) and pressure perturbations (right panels) generated by advected vortices at $R_{\rm shock} = 0.1r_s$ as function of the specific angular momentum $L$. Overall, the dependence on angular momentum $L$ is weak. The perturbations are normalized to yield a convective Mach number of $0.1$ at $r=R_\mathrm{shell}$. $L_\mathrm{max}=3{\times}10^{16}\,\mathrm{cm^2/s}$ is the maximum angular momentum considered in this work, as discussed in Section~\ref{sec:method}.}
    \label{fig:full_atRmin2}
\end{figure*}

As the vortices descend to the inner regions with higher densities, they generate strong acoustic waves. This happens because vortical motion distorts isodensity surfaces of the flow, which creates pressure perturbations that then travel as acoustic waves. The higher the density gradient, the stronger the pressure perturbations. The pressure perturbations $\delta p / p $ is of the order of density perturbations $\delta \rho / \rho$, which leads us to \citep{mueller:15} 
\begin{equation}
\label{eq:dp_est}
\frac{\delta p}{p} \sim \frac{\delta \rho}{\rho} \sim \frac{\partial \ln \rho}{\partial \ln r} 
\frac{\delta r}{r},
\end{equation}
where $\rho$ and $p$ are the mean density and pressure of the background flow. Since the radial displacements are enabled by radial velocities of the vortices, $\delta p/p$ should correlate with the radial velocities. Also, since $\delta r$ is limited by the radial size of the vortices, ${\sim} \pi R_\mathrm{shell}/m$, the vortices with large $m$ should produce weak pressure perturbations. These expectations are consistent with a more detailed calculations of \citet{abdikamalov20} for non-rotating models. Below, we explore if this holds true for rotating case. 

Figure~\ref{fig:full_atRmin2} shows the perturbation amplitudes  $\delta \upsilon_\phi/c$, $\delta \upsilon_r/c$, and $\delta p /p$ at $R_\mathrm{shock}=0.1\rso$ as a function of $L$ on the left, center, and right panels, respectively. If we compare the $\phi$ velocity to that of the vortices (shown in Fig.~\ref{fig:vort_rmin}), we notice that the acoustic waves have much larger $\phi$ velocities than the vortices. This is especially true for non- and slowly-rotating models because these vortices have smaller $\phi$ velocities. From this we can conclude that the non-radial velocity is dominated by the contribution of acoustic waves, especially in slowly rotating stars.

Figure~\ref{fig:full_atRmin2} also shows that both the velocity and pressure amplitudes decrease with increasing $m$ for most of the modes, which is consistent with the qualitative estimate (\ref{eq:dp_est}). Also, perturbations with larger $R_\mathrm{shell}$ reach smaller amplitudes at $R_\mathrm{shock}=0.1\rso$ since they undergo larger stretching due to their larger initial size. 

\begin{figure*}
    \centering
    \includegraphics[width=0.32\textwidth]{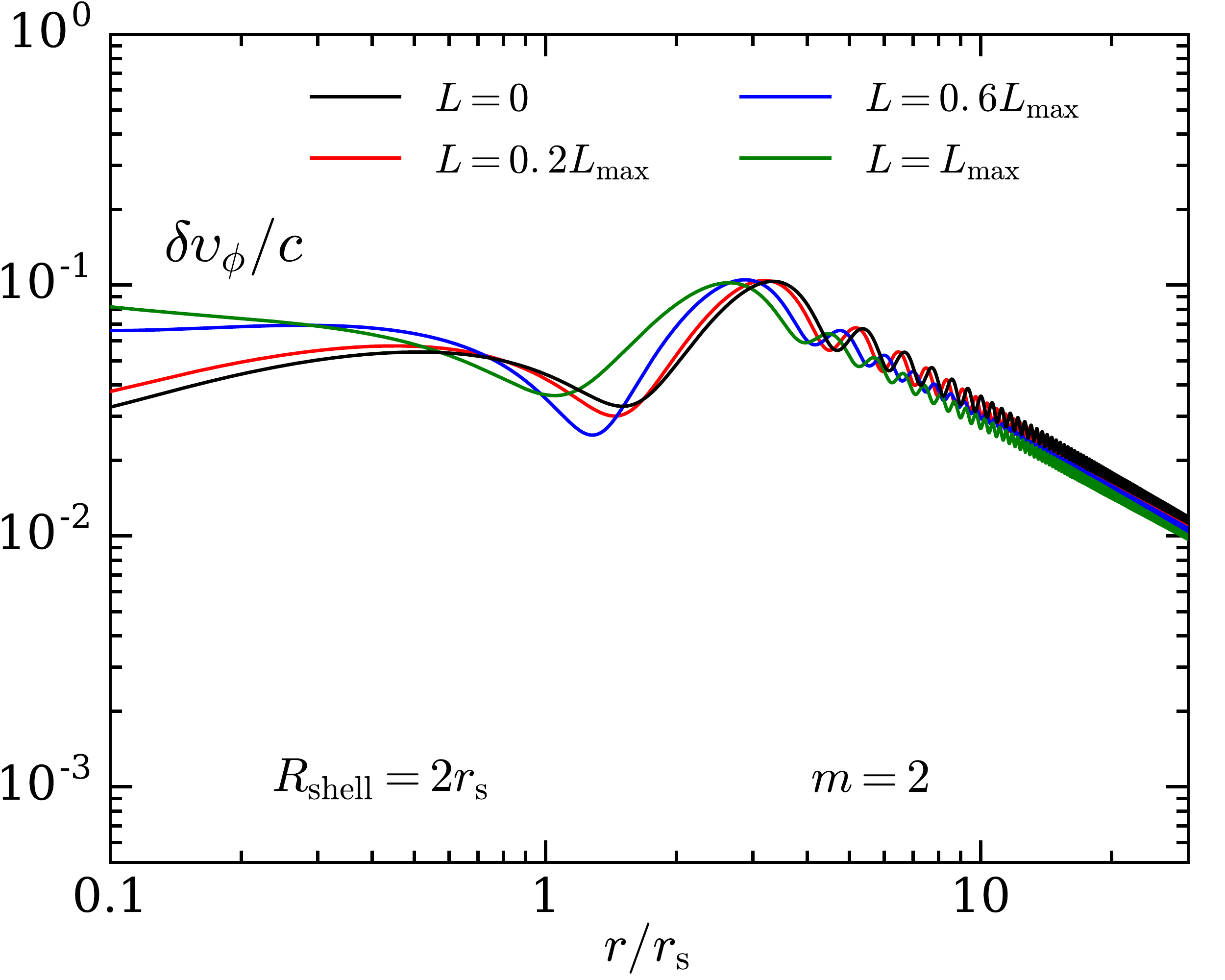}
    \includegraphics[width=0.32\textwidth]{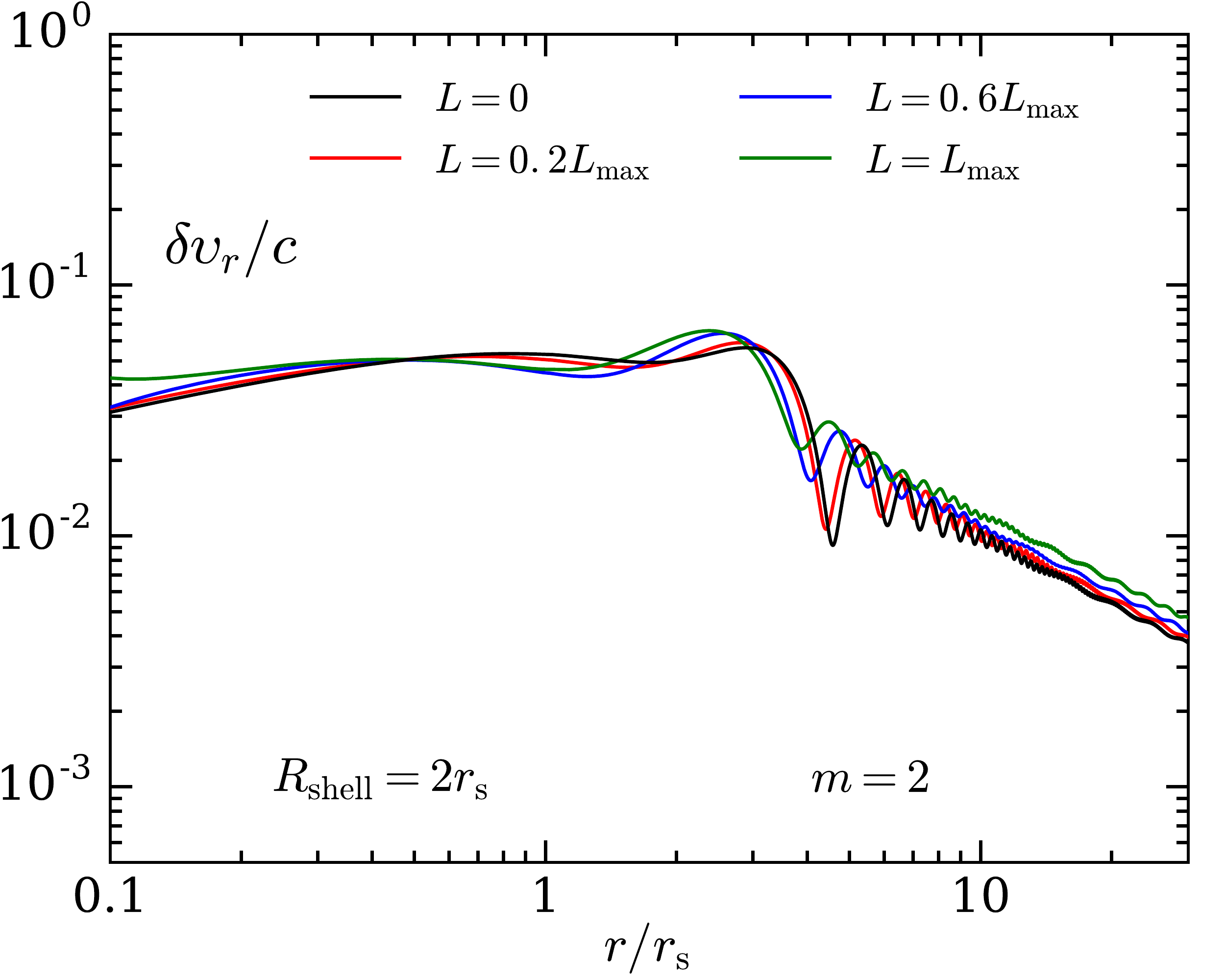}
    \includegraphics[width=0.32\textwidth]{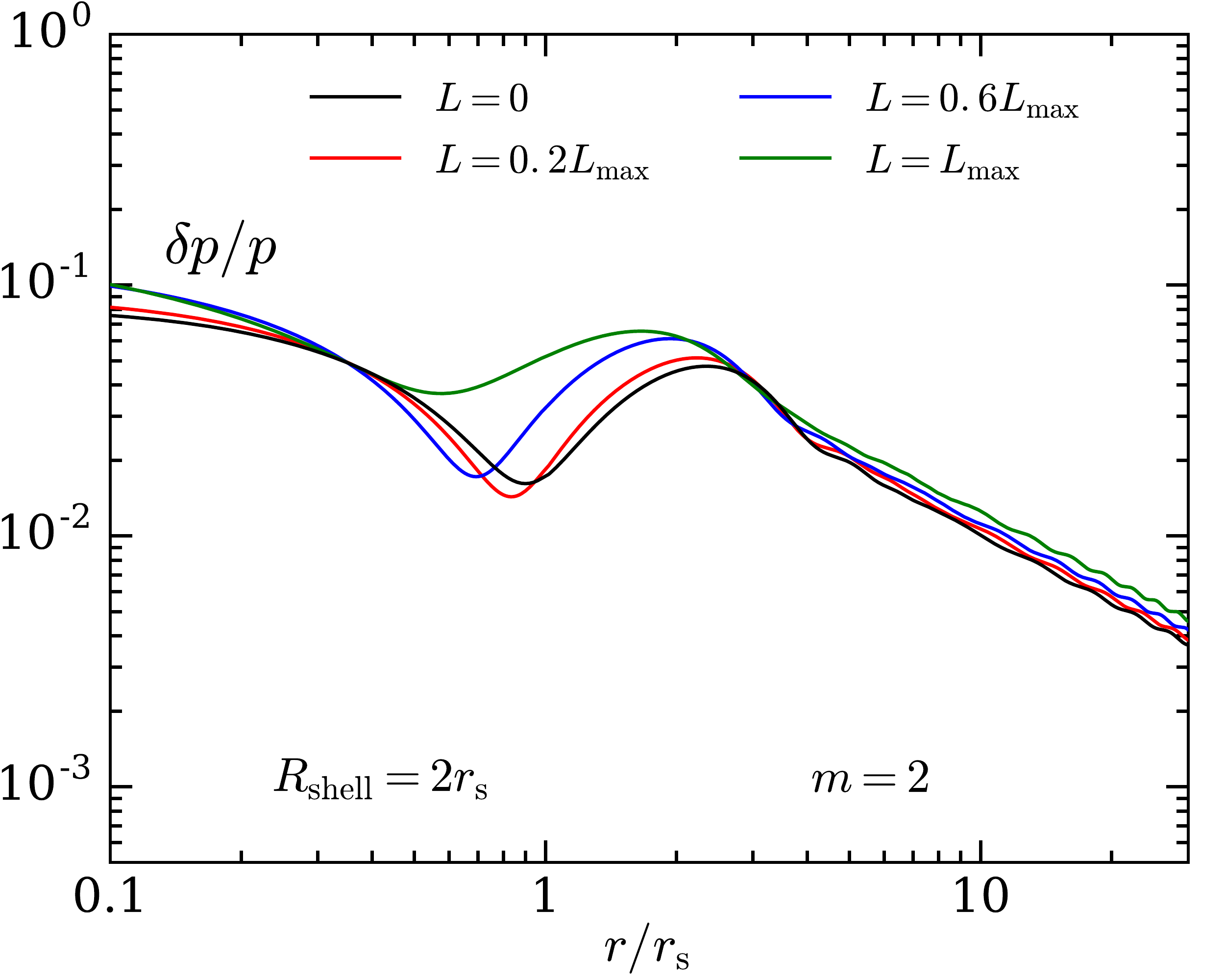}
    \caption{Velocity perturbations $\delta \upsilon_\phi/c$ (left panel) and $\delta \upsilon_r/c$ (center panel) and pressure perturbations (right panel) generated by vortices as a function of $r$ for different values of the specific angular momentum $L$ for the vortices with $m=2$ and $R_\mathrm{shell}=2\rso$. The perturbations are normalized to yield a convective Mach number of $0.1$ at $r=R_\mathrm{shell}$.}
    \label{fig:fullRepres}
\end{figure*}

Overall, Fig.~\ref{fig:full_atRmin2} shows that both velocity and pressure perturbations due to acoustic waves depend weakly on rotation. Since the emission of acoustic waves is mostly caused by the radial motion of the vortices (cf. Eq.~\ref{eq:dp_est}), the weak dependence on rotation is a result of the weak dependence of the radial velocity of the vortices on rotation that we described in the previous section. When $L$ increases from $0$ to $L_\mathrm{max}$, most modes undergo mild increase of amplitudes by ${\sim} 20\%$. This is especially true for modes with low $m$ and $R_\mathrm{shell}$ that are likely to have the strongest impact on the explosion dynamics \citep{mueller:16,kazeroni20}. Only in some rare cases, the increase can be as high as ${\sim}200\%$ for $L=L_\mathrm{max}$. Note that the vast majority of stars are expected to have rotation rates far lower than the extreme limit $L=L_\mathrm{max}$ \citep{heger:05}. For such range of $L$, the impact of rotation on the mode amplitudes is less than $\sim 10\%$. 

The same qualitative conclusion is reached when looking at the radial profiles of the velocity and pressure perturbation amplitudes, as illustrated by Fig.~\ref{fig:fullRepres} for $m=2$ perturbations with initial radius of $R_\mathrm{shell}=2\rso$. Acoustic waves at different values of $L$ are remarkably similar to each other across the entire radial domain, supporting the above observation of the weak dependence on rotation. 

\subsection{Impact of the corotation point}
\label{sec:corotation}

\begin{figure}
    \centering
    \includegraphics[width=0.45\textwidth]{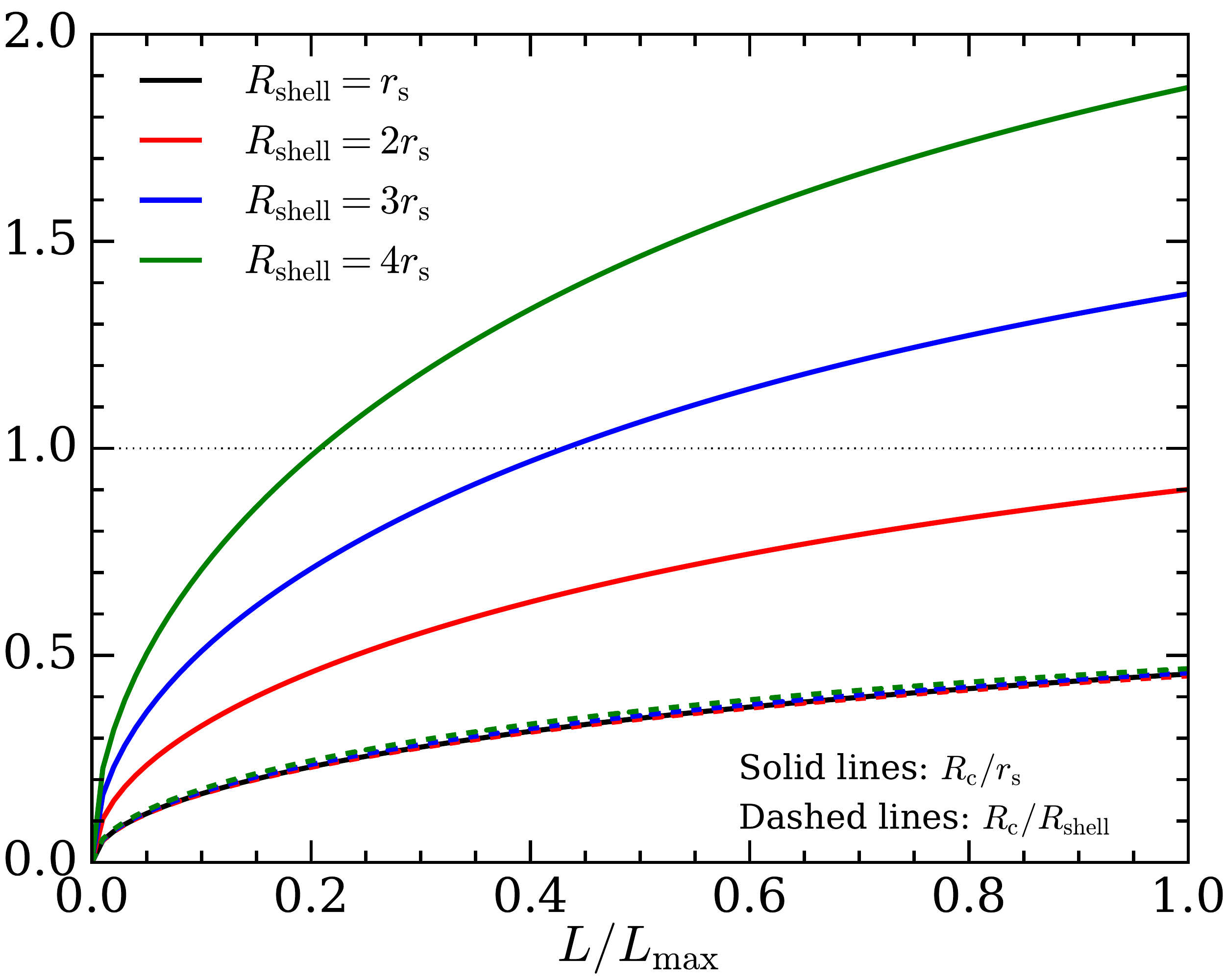}
    \caption{Corotation radius as a function of the angular momentum $L$ for perturbations originating from different initial radii $R_\mathrm{shell}$. The solid lines show the ratio of the corotation radius $\rco$ to the sonic radius $\rso$, while the dashed lines shows the ratio $\rco/R_\mathrm{shell}$. Both ratios grow with angular momentum $L$; the former is sensitive to $R_\mathrm{shell}$, while the latter is not. The horizontal dotted line represents the value of $1$.}
    \label{fig:rc}
\end{figure}

If the rotation is sufficiently rapid, the pattern speed of the vortices along the direction of rotation can match the rotation speed of the background flow.  The presence of this corotation point was shown to lead to instabilities such as the low-T/|W| instability in neutron stars \citep{watts:05, shibagaki:20}. According to \cite{yoshida:17} this instability is driven by acoustic waves trapped between the surface of the star and the corotation radius, where they over-reflect. 
The trapping of acoustic waves is unlikely in the collapsing flow because of the absence of a reflecting boundary in the outer subsonic part, and because of its supersonic character as it approaches the stalled shock. Also, since the star collapses on a dynamical free-fall timescale, the non-axisymmetric modes (if they exist and are unstable) do not have sufficient time to develop to any noticeable amplitudes. This is especially true for secular instabilities (such as the CFS instability) because their (secular) growth timescale is much longer than the timescale of stellar collapse. That said, proto-neutron stars may develop non-axisymmetric instabilities in the post-bounce phase \citep[e.g.,][for a recent review]{paschalidis:17}. 

The presence of a corotation point can still enhance the emission of acoustic waves. The phase of the source term on the right hand side of Eq.~(\ref{rdvtilde1}) is uniform in the region where $\omega'=0$, thus benefiting the generation of acoustic waves. On the other hand, our analysis in Section~\ref{sec:acoustic} shows that acoustic wave emission is not sensitive to rotation. This suggests that the presence of a corotation point should not lead to significantly stronger emission of acoustic waves. Below, we analyze the reason for the weak dependence on the presence of the corotation point.

Figure~\ref{fig:rc} shows the corotation radius as a function of the angular momentum $L$ for the vortices originating at different initial radii $R_\mathrm{shell}$. For the vortices that originate at $R_\mathrm{shell}=\rso$ and $R_\mathrm{shell}=2\rso$, the corotation point remains in the supersonic region for any $L<L_\mathrm{max}$. For $R_\mathrm{shell}=3\rso$ and $R_\mathrm{shell}=4\rso$, the corotation point is in the subsonic region for $L$ larger than $0.21L_\mathrm{max}$ and $0.43L_\mathrm{max}$, respectively. The ratio of the corotation radius $\rco$ to $R_\mathrm{shell}$ ranges from $0$ in the non-rotating case to ${\sim} 0.46$ for $L=L_\mathrm{max}$, as shown with dashed lines in Fig~\ref{fig:rc}. This ratio is not sensitive to the parameters $R_\mathrm{shell}$ and $m$. For example, the ratio changes by less than $3\%$ when the initial radius changes from $\rso$ to $4\rso$.

With this in mind, we can compare the radial size of the region where $|\omega'|$ is small. To be specific, we can define it as a region where $|\omega'|<0.25\omega$. Using Eq.~(\ref{eq:omegap}), we easily find that the radial size of this region is ${\sim} 0.25 \rco$. On the other hand, the size of the vortices is ${\sim} R_\mathrm{shell}$ for large scale modes. Since $\rco {\sim} 0.46 R_\mathrm{shell}$ for $L=L_\mathrm{max}$, we find that the region where $|\omega'| {\lesssim} 0.25 \omega$ is ${\sim} 12\%$ of the radial size of the vortex. Due to the smallness of this region compared to the size of the vortices, the presence of the corotation radius does not lead to substantially stronger emission of acoustic waves.  

\section{Conclusion}
\label{sec:conclusion}

We studied the evolution of convective vortices in collapsing massive rotating stars. We modelled convective vortices as vorticity perturbations and solved linear hydrodynamics equations on a stationary background flow. The latter was modeled using the transonic Bondi solution with rotation. The equations were solved on the equatorial plane using spherical coordinates (cf. Section~\ref{sec:method} for more details).

We find that the hydrodynamic evolution of vortices during stellar collapse is mainly governed by two effects: the acceleration of infall and speed-up of rotation during collapse. The former leads to the radial stretching of vortices, while the latter leads to deformation in the direction of rotation. The radial stretching reduces the velocity of vortices. The radial velocity component is not sensitive to rotation. As a result of the deformation in the direction of rotation, vortices acquire a strong non-radial velocity component (cf. Section~\ref{sec:vorticity} for more details). 

As vortices descend towards the central regions, they generate strong acoustic waves by distorting the iso-density surfaces. We find that the emission of acoustic waves is not sensitive to rotation. This is because the acoustic waves are mostly emitted by radial distortions and, as mentioned above, the radial velocities are not sensitive to rotation. The acoustic wave emission in rapidly rotating models is thus similar to that in non-rotating models (cf. Section~\ref{sec:acoustic} for more details). 

We analyzed the impact of the presence of a corotation point, where the pattern speed of vortices matches the rotation velocity. The Doppler-shifted frequency of the vortices vanishes at the corotation point, which favours a stronger emission of acoustic waves. However, the size of the region where the frequency is small is ${\sim}10$ times smaller than the size of the vortices. As a consequence, the corotation point does not lead to significantly stronger emission of acoustic waves (cf. Section~\ref{sec:corotation} for more details). 

While our results shed light on the evolution of convective vortices in collapsing rotating stars ahead of the accretion shock, we cannot yet make definite statements on their impact on the explosion. The explosion dynamics depends on a number of factors, including the interaction of the vortices with the shock and the post-shock flow. The impact of rotation on these processes have not yet been studied in depth. These are subjects of future studies.

We emphasize that our work is based on an idealized model, where we make targeted approximation in order to disentangle the impact of rotation from other physical processes. While such idealized models are not suitable for precise quantitative predictions, these approximations allowed us to uncover the underlying physical principles, thus complementing more detailed numerical simulations.

\section*{Acknowledgements}

The work was supported by Nazarbayev University Faculty Development Competitive Research Grant No. 090118FD5348, by the Ministry of Education of Kazakhstan's target program IRN: BR05236454 and grant AP08856149.

\section*{Data Availability}

The data underlying this article will be shared on reasonable request to the corresponding author.


\begin{thebibliography}{}
\makeatletter
\relax
\def\mn@urlcharsother{\let\do\@makeother \do\$\do\&\do\#\do\^\do\_\do\%\do\~}
\def\mn@doi{\begingroup\mn@urlcharsother \@ifnextchar [ {\mn@doi@}
  {\mn@doi@[]}}
\def\mn@doi@[#1]#2{\def\@tempa{#1}\ifx\@tempa\@empty \href
  {http://dx.doi.org/#2} {doi:#2}\else \href {http://dx.doi.org/#2} {#1}\fi
  \endgroup}
\def\mn@eprint#1#2{\mn@eprint@#1:#2::\@nil}
\def\mn@eprint@arXiv#1{\href {http://arxiv.org/abs/#1} {{\tt arXiv:#1}}}
\def\mn@eprint@dblp#1{\href {http://dblp.uni-trier.de/rec/bibtex/#1.xml}
  {dblp:#1}}
\def\mn@eprint@#1:#2:#3:#4\@nil{\def\@tempa {#1}\def\@tempb {#2}\def\@tempc
  {#3}\ifx \@tempc \@empty \let \@tempc \@tempb \let \@tempb \@tempa \fi \ifx
  \@tempb \@empty \def\@tempb {arXiv}\fi \@ifundefined
  {mn@eprint@\@tempb}{\@tempb:\@tempc}{\expandafter \expandafter \csname
  mn@eprint@\@tempb\endcsname \expandafter{\@tempc}}}

\bibitem[\protect\citeauthoryear{{Abdikamalov} \& {Foglizzo}}{{Abdikamalov} \&
  {Foglizzo}}{2020}]{abdikamalov20}
{Abdikamalov} E.,  {Foglizzo} T.,  2020, \mn@doi [\mnras]
  {10.1093/mnras/staa533}, \href
  {https://ui.adsabs.harvard.edu/abs/2020MNRAS.493.3496A} {493, 3496}

\bibitem[\protect\citeauthoryear{{Abdikamalov}, {Zhaksylykov}, {Radice}  \&
  {Berdibek}}{{Abdikamalov} et~al.}{2016}]{abdikamalov:16}
{Abdikamalov} E.,  {Zhaksylykov} A.,  {Radice} D.,   {Berdibek} S.,  2016,
  \mn@doi [\mnras] {10.1093/mnras/stw1604}, \href
  {http://adsabs.harvard.edu/abs/2016MNRAS.461.3864A} {461, 3864}

\bibitem[\protect\citeauthoryear{Abdikamalov, Huete, Nussupbekov  \&
  Berdibek}{Abdikamalov et~al.}{2018}]{abdikamalov18}
Abdikamalov E.,  Huete C.,  Nussupbekov A.,   Berdibek S.,  2018, \mn@doi
  [Particles] {10.3390/particles1010007}, 1, 7

\bibitem[\protect\citeauthoryear{{Akiyama}, {Wheeler}, {Meier}  \&
  {Lichtenstadt}}{{Akiyama} et~al.}{2003}]{akiyama:03}
{Akiyama} S.,  {Wheeler} J.~C.,  {Meier} D.~L.,   {Lichtenstadt} I.,  2003,
  \mn@doi [\apj] {10.1086/344135}, \href
  {http://adsabs.harvard.edu/abs/2003ApJ...584..954A} {584, 954}

\bibitem[\protect\citeauthoryear{{Arnett}}{{Arnett}}{1996}]{arnett:96}
{Arnett} D.,  1996, {Supernovae and Nucleosynthesis}.
Princeton University Press, Princeton NJ, United States

\bibitem[\protect\citeauthoryear{{Bisnovatyi-Kogan}, {Popov}  \&
  {Samokhin}}{{Bisnovatyi-Kogan} et~al.}{1976}]{bisno:76}
{Bisnovatyi-Kogan} G.~S.,  {Popov} I.~P.,   {Samokhin} A.~A.,  1976, \mn@doi
  [\apss] {10.1007/BF00646184}, \href
  {http://adsabs.harvard.edu/abs/1976Ap%26SS..41..287B} {41, 287}

\bibitem[\protect\citeauthoryear{{Buras}, {Janka}, {Rampp}  \&
  {Kifonidis}}{{Buras} et~al.}{2006}]{buras:06b}
{Buras} R.,  {Janka} H.-T.,  {Rampp} M.,   {Kifonidis} K.,  2006, \aap, 457,
  281

\bibitem[\protect\citeauthoryear{{Burrows}, {Dessart}, {Livne}, {Ott}  \&
  {Murphy}}{{Burrows} et~al.}{2007}]{burrows:07b}
{Burrows} A.,  {Dessart} L.,  {Livne} E.,  {Ott} C.~D.,   {Murphy} J.,  2007,
  \mn@doi [\apj] {10.1086/519161}, \href
  {http://adsabs.harvard.edu/abs/2007ApJ...664..416B} {664, 416}

\bibitem[\protect\citeauthoryear{{Cantiello}, {Mankovich}, {Bildsten},
  {Christensen-Dalsgaard}  \& {Paxton}}{{Cantiello}
  et~al.}{2014}]{cantiello:14}
{Cantiello} M.,  {Mankovich} C.,  {Bildsten} L.,  {Christensen-Dalsgaard} J.,
  {Paxton} B.,  2014, \mn@doi [\apj] {10.1088/0004-637X/788/1/93}, \href
  {https://ui.adsabs.harvard.edu/abs/2014ApJ...788...93C} {788, 93}

\bibitem[\protect\citeauthoryear{{Chatzopoulos}, {Graziani}  \&
  {Couch}}{{Chatzopoulos} et~al.}{2014}]{chatzopoulos:14}
{Chatzopoulos} E.,  {Graziani} C.,   {Couch} S.~M.,  2014, \mn@doi [\apj]
  {10.1088/0004-637X/795/1/92}, \href
  {http://adsabs.harvard.edu/abs/2014ApJ...795...92C} {795, 92}

\bibitem[\protect\citeauthoryear{{Collins}, {M{\"u}ller}  \& {Heger}}{{Collins}
  et~al.}{2018}]{collins:18}
{Collins} C.,  {M{\"u}ller} B.,   {Heger} A.,  2018, \mn@doi [\mnras]
  {10.1093/mnras/stx2470}, \href
  {http://adsabs.harvard.edu/abs/2018MNRAS.473.1695C} {473, 1695}

\bibitem[\protect\citeauthoryear{{Couch} \& {Ott}}{{Couch} \&
  {Ott}}{2013}]{couch:13d}
{Couch} S.~M.,  {Ott} C.~D.,  2013, \apjl, \href
  {http://adsabs.harvard.edu/abs/2013arXiv1309.2632C} {778, L7}

\bibitem[\protect\citeauthoryear{{Couch}, {Chatzopoulos}, {Arnett}  \&
  {Timmes}}{{Couch} et~al.}{2015}]{couch:15b}
{Couch} S.~M.,  {Chatzopoulos} E.,  {Arnett} W.~D.,   {Timmes} F.~X.,  2015,
  \mn@doi [\apjl] {10.1088/2041-8205/808/1/L21}, \href
  {http://adsabs.harvard.edu/abs/2015ApJ...808L..21C} {808, L21}

\bibitem[\protect\citeauthoryear{{Deheuvels} et~al.,}{{Deheuvels}
  et~al.}{2014}]{deheuvels:14}
{Deheuvels} S.,  et~al., 2014, \mn@doi [\aap] {10.1051/0004-6361/201322779},
  \href {https://ui.adsabs.harvard.edu/abs/2014A&A...564A..27D} {564, A27}

\bibitem[\protect\citeauthoryear{{Endeve}, {Cardall}, {Budiardja}  \&
  {Mezzacappa}}{{Endeve} et~al.}{2010}]{endeve:10}
{Endeve} E.,  {Cardall} C.~Y.,  {Budiardja} R.~D.,   {Mezzacappa} A.,  2010,
  \apj, \href {http://adsabs.harvard.edu/abs/2010ApJ...713.1219E} {713, 1219}

\bibitem[\protect\citeauthoryear{{Endeve}, {Cardall}, {Budiardja}, {Beck},
  {Bejnood}, {Toedte}, {Mezzacappa}  \& {Blondin}}{{Endeve}
  et~al.}{2012}]{endeve:12}
{Endeve} E.,  {Cardall} C.~Y.,  {Budiardja} R.~D.,  {Beck} S.~W.,  {Bejnood}
  A.,  {Toedte} R.~J.,  {Mezzacappa} A.,   {Blondin} J.~M.,  2012, \mn@doi
  [\apj] {10.1088/0004-637X/751/1/26}, \href
  {http://adsabs.harvard.edu/abs/2012ApJ...751...26E} {751, 26}

\bibitem[\protect\citeauthoryear{{Foglizzo}}{{Foglizzo}}{2001}]{foglizzo:01}
{Foglizzo} T.,  2001, \mn@doi [\aap] {10.1051/0004-6361:20000506}, \href
  {http://adsabs.harvard.edu/abs/2001A%26A...368..311F} {368, 311}

\bibitem[\protect\citeauthoryear{{Fujisawa}, {Okawa}, {Yamamoto}  \&
  {Yamada}}{{Fujisawa} et~al.}{2019}]{fujisawa19}
{Fujisawa} K.,  {Okawa} H.,  {Yamamoto} Y.,   {Yamada} S.,  2019, \mn@doi
  [\apj] {10.3847/1538-4357/aaffdd}, \href
  {https://ui.adsabs.harvard.edu/abs/2019ApJ...872..155F} {872, 155}

\bibitem[\protect\citeauthoryear{{Heger}, {Woosley}  \& {Spruit}}{{Heger}
  et~al.}{2005}]{heger:05}
{Heger} A.,  {Woosley} S.~E.,   {Spruit} H.~C.,  2005, \apj, 626, 350

\bibitem[\protect\citeauthoryear{Huete \& Abdikamalov}{Huete \&
  Abdikamalov}{2019}]{huete:19}
Huete C.,  Abdikamalov E.,  2019, \mn@doi [Physica Scripta]
  {10.1088/1402-4896/ab0228}, 94, 094002

\bibitem[\protect\citeauthoryear{{Huete}, {Abdikamalov}  \& {Radice}}{{Huete}
  et~al.}{2018}]{huete:18}
{Huete} C.,  {Abdikamalov} E.,   {Radice} D.,  2018, \mn@doi [\mnras]
  {10.1093/mnras/stx3360}, \href
  {http://adsabs.harvard.edu/abs/2018MNRAS.475.3305H} {475, 3305}

\bibitem[\protect\citeauthoryear{{Janka}}{{Janka}}{2012}]{janka:12a}
{Janka} H.-T.,  2012, \mn@doi [Ann. Rev. Nuc. Par. Sci.]
  {10.1146/annurev-nucl-102711-094901}, \href
  {http://adsabs.harvard.edu/abs/2012ARNPS..62..407J} {62, 407}

\bibitem[\protect\citeauthoryear{{Kazeroni} \& {Abdikamalov}}{{Kazeroni} \&
  {Abdikamalov}}{2020}]{kazeroni20}
{Kazeroni} R.,  {Abdikamalov} E.,  2020, \mn@doi [\mnras]
  {10.1093/mnras/staa944}, \href
  {https://ui.adsabs.harvard.edu/abs/2020MNRAS.494.5360K} {494, 5360}

\bibitem[\protect\citeauthoryear{{Kovalenko} \& {Eremin}}{{Kovalenko} \&
  {Eremin}}{1998}]{kovalenko:98}
{Kovalenko} I.~G.,  {Eremin} M.~A.,  1998, \mn@doi [\mnras]
  {10.1046/j.1365-8711.1998.01667.x}, \href
  {http://adsabs.harvard.edu/abs/1998MNRAS.298..861K} {298, 861}

\bibitem[\protect\citeauthoryear{{Kuroda}, {Arcones}, {Takiwaki}  \&
  {Kotake}}{{Kuroda} et~al.}{2020}]{kuroda20}
{Kuroda} T.,  {Arcones} A.,  {Takiwaki} T.,   {Kotake} K.,  2020, \mn@doi
  [\apj] {10.3847/1538-4357/ab9308}, \href
  {https://ui.adsabs.harvard.edu/abs/2020ApJ...896..102K} {896, 102}

\bibitem[\protect\citeauthoryear{{Lai} \& {Goldreich}}{{Lai} \&
  {Goldreich}}{2000}]{lai:00}
{Lai} D.,  {Goldreich} P.,  2000, \mn@doi [\apj] {10.1086/308821}, \href
  {http://adsabs.harvard.edu/abs/2000ApJ...535..402L} {535, 402}

\bibitem[\protect\citeauthoryear{{Landau} \& {Lifshitz}}{{Landau} \&
  {Lifshitz}}{1959}]{landau:59}
{Landau} L.~D.,  {Lifshitz} E.~M.,  1959, {Fluid Mechanics, 2nd edition}.
Butterworth-Heinemann, Oxford, UK

\bibitem[\protect\citeauthoryear{{Meier}, {Epstein}, {Arnett}  \&
  {Schramm}}{{Meier} et~al.}{1976}]{meier:76}
{Meier} D.~L.,  {Epstein} R.~I.,  {Arnett} W.~D.,   {Schramm} D.~N.,  1976,
  \mn@doi [\apj] {10.1086/154235}, \href
  {http://adsabs.harvard.edu/abs/1976ApJ...204..869M} {204, 869}

\bibitem[\protect\citeauthoryear{{Mosser} et~al.,}{{Mosser}
  et~al.}{2012}]{mosser:12}
{Mosser} B.,  et~al., 2012, \mn@doi [\aap] {10.1051/0004-6361/201220106}, \href
  {https://ui.adsabs.harvard.edu/abs/2012A&A...548A..10M} {548, A10}

\bibitem[\protect\citeauthoryear{{M{\"o}sta} et~al.,}{{M{\"o}sta}
  et~al.}{2014}]{moesta:14b}
{M{\"o}sta} P.,  et~al., 2014, \mn@doi [\apjl] {10.1088/2041-8205/785/2/L29},
  \href {http://adsabs.harvard.edu/abs/2014ApJ...785L..29M} {785, L29}

\bibitem[\protect\citeauthoryear{{M{\"u}ller}}{{M{\"u}ller}}{2020}]{mueller:20review}
{M{\"u}ller} B.,  2020, \mn@doi [Living Reviews in Computational Astrophysics]
  {10.1007/s41115-020-0008-5}, \href
  {https://ui.adsabs.harvard.edu/abs/2020LRCA....6....3M} {6, 3}

\bibitem[\protect\citeauthoryear{{M{\"u}ller} \& {Janka}}{{M{\"u}ller} \&
  {Janka}}{2015}]{mueller:15}
{M{\"u}ller} B.,  {Janka} H.-T.,  2015, \mn@doi [\mnras]
  {10.1093/mnras/stv101}, \href
  {http://adsabs.harvard.edu/abs/2015MNRAS.448.2141M} {448, 2141}

\bibitem[\protect\citeauthoryear{{M{\"u}ller} \& {Varma}}{{M{\"u}ller} \&
  {Varma}}{2020}]{mueller20b}
{M{\"u}ller} B.,  {Varma} V.,  2020, \mn@doi [\mnras] {10.1093/mnrasl/slaa137},
  \href {https://ui.adsabs.harvard.edu/abs/2020MNRAS.498L.109M} {498, L109}

\bibitem[\protect\citeauthoryear{{M{\"u}ller}, {Viallet}, {Heger}  \&
  {Janka}}{{M{\"u}ller} et~al.}{2016}]{mueller:16}
{M{\"u}ller} B.,  {Viallet} M.,  {Heger} A.,   {Janka} H.-T.,  2016, \mn@doi
  [\apj] {10.3847/1538-4357/833/1/124}, \href
  {http://adsabs.harvard.edu/abs/2016ApJ...833..124M} {833, 124}

\bibitem[\protect\citeauthoryear{{M{\"u}ller}, {Melson}, {Heger}  \&
  {Janka}}{{M{\"u}ller} et~al.}{2017}]{mueller:17}
{M{\"u}ller} B.,  {Melson} T.,  {Heger} A.,   {Janka} H.-T.,  2017, \mn@doi
  [\mnras] {10.1093/mnras/stx1962}, \href
  {http://adsabs.harvard.edu/abs/2017MNRAS.472..491M} {472, 491}

\bibitem[\protect\citeauthoryear{{Nagakura}, {Takahashi}  \&
  {Yamamoto}}{{Nagakura} et~al.}{2019}]{nagakura:19}
{Nagakura} H.,  {Takahashi} K.,   {Yamamoto} Y.,  2019, \mn@doi [\mnras]
  {10.1093/mnras/sty3114}, \href
  {https://ui.adsabs.harvard.edu/abs/2019MNRAS.483..208N} {483, 208}

\bibitem[\protect\citeauthoryear{{Nagakura}, {Burrows}, {Radice}  \&
  {Vartanyan}}{{Nagakura} et~al.}{2020}]{nagakura20}
{Nagakura} H.,  {Burrows} A.,  {Radice} D.,   {Vartanyan} D.,  2020, \mn@doi
  [\mnras] {10.1093/mnras/staa261}, \href
  {https://ui.adsabs.harvard.edu/abs/2020MNRAS.492.5764N} {492, 5764}

\bibitem[\protect\citeauthoryear{{Nakamura}, {Kuroda}, {Takiwaki}  \&
  {Kotake}}{{Nakamura} et~al.}{2014}]{nakamura:14}
{Nakamura} K.,  {Kuroda} T.,  {Takiwaki} T.,   {Kotake} K.,  2014, \mn@doi
  [\apj] {10.1088/0004-637X/793/1/45}, \href
  {http://adsabs.harvard.edu/abs/2014ApJ...793...45N} {793, 45}

\bibitem[\protect\citeauthoryear{{Obergaulinger} \& {Aloy}}{{Obergaulinger} \&
  {Aloy}}{2020}]{obergaulinger20a}
{Obergaulinger} M.,  {Aloy} M.~{\'A}.,  2020, \mn@doi [\mnras]
  {10.1093/mnras/staa096}, \href
  {https://ui.adsabs.harvard.edu/abs/2020MNRAS.492.4613O} {492, 4613}

\bibitem[\protect\citeauthoryear{{Paschalidis} \& {Stergioulas}}{{Paschalidis}
  \& {Stergioulas}}{2017}]{paschalidis:17}
{Paschalidis} V.,  {Stergioulas} N.,  2017, \mn@doi [Living Reviews in
  Relativity] {10.1007/s41114-017-0008-x}, \href
  {https://ui.adsabs.harvard.edu/abs/2017LRR....20....7P} {20, 7}

\bibitem[\protect\citeauthoryear{{Popov} \& {Turolla}}{{Popov} \&
  {Turolla}}{2012}]{popov:12}
{Popov} S.~B.,  {Turolla} R.,  2012, \mn@doi [\apss]
  {10.1007/s10509-012-1100-z}, \href
  {https://ui.adsabs.harvard.edu/abs/2012Ap&SS.341..457P} {341, 457}

\bibitem[\protect\citeauthoryear{{Raynaud}, {Guilet}, {Janka}  \&
  {Gastine}}{{Raynaud} et~al.}{2020}]{raynaud20}
{Raynaud} R.,  {Guilet} J.,  {Janka} H.-T.,   {Gastine} T.,  2020, \mn@doi
  [Science Advances] {10.1126/sciadv.aay2732}, \href
  {https://ui.adsabs.harvard.edu/abs/2020SciA....6.2732R} {6, eaay2732}

\bibitem[\protect\citeauthoryear{{Shibagaki}, {Kuroda}, {Kotake}  \&
  {Takiwaki}}{{Shibagaki} et~al.}{2020}]{shibagaki:20}
{Shibagaki} S.,  {Kuroda} T.,  {Kotake} K.,   {Takiwaki} T.,  2020, \mn@doi
  [\mnras] {10.1093/mnrasl/slaa021}, \href
  {https://ui.adsabs.harvard.edu/abs/2020MNRAS.493L.138S} {493, L138}

\bibitem[\protect\citeauthoryear{{Summa}, {Janka}, {Melson}  \&
  {Marek}}{{Summa} et~al.}{2018}]{summa:18}
{Summa} A.,  {Janka} H.-T.,  {Melson} T.,   {Marek} A.,  2018, \mn@doi [\apj]
  {10.3847/1538-4357/aa9ce8}, \href
  {https://ui.adsabs.harvard.edu/abs/2018ApJ...852...28S} {852, 28}

\bibitem[\protect\citeauthoryear{{Takahashi} \& {Yamada}}{{Takahashi} \&
  {Yamada}}{2014}]{takahashi:14}
{Takahashi} K.,  {Yamada} S.,  2014, \mn@doi [\apj]
  {10.1088/0004-637X/794/2/162}, \href
  {http://adsabs.harvard.edu/abs/2014ApJ...794..162T} {794, 162}

\bibitem[\protect\citeauthoryear{{Takahashi}, {Iwakami}, {Yamamoto}  \&
  {Yamada}}{{Takahashi} et~al.}{2016}]{takahashi:16}
{Takahashi} K.,  {Iwakami} W.,  {Yamamoto} Y.,   {Yamada} S.,  2016, \mn@doi
  [\apj] {10.3847/0004-637X/831/1/75}, \href
  {http://adsabs.harvard.edu/abs/2016ApJ...831...75T} {831, 75}

\bibitem[\protect\citeauthoryear{Tassoul}{Tassoul}{1978}]{tassoul:78}
Tassoul J.-L.,  1978, Theory of Rotating Stars.
Princeton University Press, Princeton U. S. A.

\bibitem[\protect\citeauthoryear{{Watts}, {Andersson}  \& {Jones}}{{Watts}
  et~al.}{2005}]{watts:05}
{Watts} A.~L.,  {Andersson} N.,   {Jones} D.~I.,  2005, \mn@doi [\apjl]
  {10.1086/427653}, \href
  {https://ui.adsabs.harvard.edu/abs/2005ApJ...618L..37W} {618, L37}

\bibitem[\protect\citeauthoryear{{Winteler}, {K{\"a}ppeli}, {Perego},
  {Arcones}, {Vasset}, {Nishimura}, {Liebend{\"o}rfer}  \&
  {Thielemann}}{{Winteler} et~al.}{2012}]{winteler12}
{Winteler} C.,  {K{\"a}ppeli} R.,  {Perego} A.,  {Arcones} A.,  {Vasset} N.,
  {Nishimura} N.,  {Liebend{\"o}rfer} M.,   {Thielemann} F.~K.,  2012, \mn@doi
  [\apjl] {10.1088/2041-8205/750/1/L22}, \href
  {https://ui.adsabs.harvard.edu/abs/2012ApJ...750L..22W} {750, L22}

\bibitem[\protect\citeauthoryear{{Woosley} \& {Heger}}{{Woosley} \&
  {Heger}}{2006}]{woosley:06}
{Woosley} S.~E.,  {Heger} A.,  2006, \apj, 637, 914

\bibitem[\protect\citeauthoryear{{Yadav}, {M{\"u}ller}, {Janka}, {Melson}  \&
  {Heger}}{{Yadav} et~al.}{2020}]{yadav20}
{Yadav} N.,  {M{\"u}ller} B.,  {Janka} H.~T.,  {Melson} T.,   {Heger} A.,
  2020, \mn@doi [\apj] {10.3847/1538-4357/ab66bb}, \href
  {https://ui.adsabs.harvard.edu/abs/2020ApJ...890...94Y} {890, 94}

\bibitem[\protect\citeauthoryear{{Yamamoto}, {Fujimoto}, {Nagakura}  \&
  {Yamada}}{{Yamamoto} et~al.}{2013}]{yamamoto:13}
{Yamamoto} Y.,  {Fujimoto} S.-i.,  {Nagakura} H.,   {Yamada} S.,  2013, \mn@doi
  [\apj] {10.1088/0004-637X/771/1/27}, \href
  {https://ui.adsabs.harvard.edu/abs/2013ApJ...771...27Y} {771, 27}

\bibitem[\protect\citeauthoryear{{Yamasaki} \& {Foglizzo}}{{Yamasaki} \&
  {Foglizzo}}{2008}]{yamasaki:08}
{Yamasaki} T.,  {Foglizzo} T.,  2008, \apj, 679, 607

\bibitem[\protect\citeauthoryear{{Yoshida} \& {Saijo}}{{Yoshida} \&
  {Saijo}}{2017}]{yoshida:17}
{Yoshida} S.,  {Saijo} M.,  2017, \mn@doi [\mnras] {10.1093/mnras/stw3064},
  \href {https://ui.adsabs.harvard.edu/abs/2017MNRAS.466..600Y} {466, 600}

\bibitem[\protect\citeauthoryear{Yoshida, Takiwaki, Kotake, Takahashi, Nakamura
   \& Umeda}{Yoshida et~al.}{2019}]{yoshida19}
Yoshida T.,  Takiwaki T.,  Kotake K.,  Takahashi K.,  Nakamura K.,   Umeda H.,
  2019, \mn@doi [The Astrophysical Journal] {10.3847/1538-4357/ab2b9d}, 881, 16

\makeatother
\end{thebibliography}

\appendix

\section{Mathematical Formalism}
\label{sec:formalism}

The set of equations describing the flow is the following
\begin{eqnarray}
{\p \rho\over \p t}+\boldsymbol{\nabla}\cdot\rho \boldsymbol{\upsilon} &=&0,\\
{\p  \boldsymbol{\upsilon}\over\p t}+\boldsymbol{w} \times \boldsymbol{\upsilon} +\boldsymbol{\nabla} \left({v^2\over 2}\!+\!{c^2\over\gamma-1}\!+\!\Phi_0\right)&=&{c^2\over\gamma}\boldsymbol{\nabla} S,\\
{\p S\over \p t}+\boldsymbol{\upsilon}\cdot\boldsymbol{\nabla} S&=&0,
\end{eqnarray}
where $\rho$ is density, $\upsilon$ is velocity, $\boldsymbol{w} = \nabla \times \boldsymbol{\upsilon}$ is vorticity of the flow, $c$ is the speed of sound, $S$ is entropy, and $\Phi_0$ is the gravitational potential. We write the equations in spherical coordinates $(r,\theta,\varphi)$ restricted to the equatorial plane $\theta=\pi/2$, assuming zero derivatives in the $\theta$-direction. We use the adiabatic approximation. The gravitational potential is given by $\Phi_0\equiv -GM/r$, where $M$ is the neutron star mass. The stellar matter is modeled with an ideal gas equation of state with adiabatic index $\gamma=4/3$, which is a good approximation for the radiaion-dominated stellar gas \citep[e.g.,][]{arnett:96}.

\section{Stationary transonic adiabatic solution}
\label{sec:stationary_solution}

The stationary background flow is axisymmetric with constant entropy and specific angular momentum. On the equatorial plane, the flow is given by the stationary equations of continuity, Euler, and entropy:
\begin{eqnarray}
4 \pi r^2\rho \upsilon_r&=&\dot M ,\\ 
v_\varphi&\equiv&{L\over r} ,\\
w_\theta&=&0,\\
{\p S\over\p r}&=&0,\\
{\p\over\p r}\left({\upsilon_r^2\over 2}+{L^2\over 2r^2}+{c^2\over\gamma-1}+\Phi_0\right)&=&
0,
\end{eqnarray}
where condition of vanishing vorticity $w_\theta=0$ stems from the axisymmetry of the stationary background flow. Here, $\upsilon_r$ and $\upsilon_\phi$ are the $r$ and $\phi$ components of the velocity, while $\dot M$ is the mass accretion rate.

The gas at rest at infinity is defined by three variables: its specific angular momentum $L$, its sound speed $c_\infty$ and entropy $S_\infty$, or its density $\rho_\infty$ and pressure $P_\infty$ with the following relations:
\begin{eqnarray}
P_\infty&=&\rho_\infty {c_\infty^2\over\gamma},\\
S_\infty&=&{1\over\gamma-1}{P_\infty\over\rho_\infty}.
\end{eqnarray}
The maximum mass accretion rate of a solution accreting on a point-like mass $M$ corresponds to the transonic solution such that the Mach number reaches unity at the sonic radius $\rso$,
\begin{eqnarray}
{\upsilon_r^2\over2}+{c^2\over\gamma-1}&=&{GM\over r}-{L^2\over 2r^2}+{c_\infty^2\over\gamma-1}, \label{Bernoulli}\\
r^2\M c^{\gamma+1\over \gamma-1}&=&\rso^2\cs^{\gamma+1\over\gamma-1},
\end{eqnarray}
where $\M$ is the Mach number of the radial infall and $\cs$ is the speed of sound at the sonic point $\rso$. Differentiating these equations:
\begin{eqnarray}
(c^2-\upsilon_r^2){\dot \upsilon_r\over \upsilon_r}={GM\over r^2}-{L^2\over r^3}-{c^2\over r}.
\end{eqnarray}
From this we can obtain the speed of sound at the sonic radius:
\begin{eqnarray}
\cs^2={GM\over 2\rso}-{L^2\over 2\rso^2}.\label{cson}
\end{eqnarray}
The energy density of the transonic solution is defined by the regularity condition:
\begin{eqnarray}
c_\infty^2 &=& \frac{5-3\gamma}{4} \frac{GM}{\rso} - \frac{3-\gamma}{4} \frac{L^2}{\rso^2}.
\end{eqnarray}
The position of the sonic point depends quadratically on the rotation rate:
\begin{eqnarray}
\rso = {5-3\gamma \over 8}{GM \over c_\infty^2} Q{\left( \frac{Lc_\infty}{GM} \right)}, \label{rson}
\end{eqnarray}
where
\begin{eqnarray}
    Q(x)=1+\left\lbrack 1-(3-\gamma)\left({4x\over 5-3\gamma}\right)^2\right\rbrack^{1\over 2}.
\end{eqnarray}
From the adiabatic hypothesis,
\begin{eqnarray}
{\rho_{\rm s}\over\rho_{\infty}}=\left({\cs\over c_{\infty}}\right)^{2\over\gamma-1},
\end{eqnarray}
where $\rho_{\rm s}$ is the density at sonic radius $\rso$.
The mass accretion rate $\dot M_{\rm trans}$ of the transonic solution is thus determined by $\rho_{\infty}$, $c_{\infty}$, $M$ and $L$, using Eqs.~(\ref{cson}) and (\ref{rson}):
\begin{eqnarray}
\dot M_{\rm trans}&\equiv& 4\pi \rso^2\rho_{\rm s}\cs,\nonumber\\
&=&4\pi \rso^2\rho_{\infty}c_{\infty}\left({\cs\over c_{\infty}}\right)^{\gamma+1\over\gamma-1},\nonumber \\
&=&\left({5-3\gamma\over 4}\right)^2{\rho_\infty G^2M^2\over c_\infty^3}F\left({Lc_\infty\over GM}\right),
\end{eqnarray}
where
\begin{eqnarray}
F(x)&=&4\pi\left({2\over 5-3\gamma}\right)^{\gamma+1\over 2(\gamma-1)} \nonumber \\
&\times &
Q^2(x)\!
\left[ 
Q(x)-{8x^2\over 5-3\gamma}\over Q(x)\!-\!(3\!-\!\gamma){8x^2\over (5-3\gamma)^2}\right]^{\gamma+1\over 2(\gamma-1)}\!.
\end{eqnarray}
The radial profile of the Mach number of the transonic solution depends only on $c_{\infty}$, $M$ and $L$, as obtained by solving the stationary equations:
\begin{eqnarray}
 \left({\M^2\over 2}+{1\over\gamma-1}\right)\left({\rso\over r}\right)^{4(\gamma-1)\over\gamma+1}{1\over \M^{2(\gamma-1)\over\gamma+1}} =
{GM\over r\cs^2} \nonumber \\
-{L^2\over 2r^2\cs^2}+{1\over\gamma-1}{c_\infty^2\over \cs^2}.
\label{implicit}
\end{eqnarray}
with
\begin{eqnarray}
\label{eq:c}
{c\over \cs}&=&\left({\rso\over r}\right)^{2(\gamma-1)\over\gamma+1}{1\over \M^{\gamma-1\over\gamma+1}},\\
\upsilon_r&=&-\M c \label{eq:v}
\end{eqnarray}
We can simplify Eqs.~(\ref{implicit})-(\ref{eq:c}) further if we measure distance, velocity, and angular momentum in units of radius $r_\mathrm{B}=GM/c_\infty^2$, $c_\infty$, and $(GMr_\mathrm{B})^{1/2}$, respectively:
\begin{eqnarray}
 {\upsilon_r^2\over 2}+{c^2\over\gamma-1} &=& {1\over r} 
-{L^2\over 2r^2}+{1\over\gamma-1}. \label{eq:bondi1} \\
r^2 \upsilon_r c^{\frac{2}{\gamma-1}} &=& \rso^2 c^{\frac{\gamma+1}{\gamma-1}}. \label{eq:bondi2}
\end{eqnarray}
The scale of this equation can be fixed by selecting the values of $L$ and $\rso$. The speed of sound $\cs$ is related to $L$ and $\rso$ via Eq.~(\ref{cson}), which, using these units, can be re-written as 
\begin{eqnarray}
\cs^2={1\over 2\rso}-{L^2\over 2\rso^2}, \label{cson2}
\end{eqnarray}
while Eq.~(\ref{rson}) becomes 
\begin{eqnarray}
\rso = {5-3\gamma \over 8} Q{\left(L \right)}, \label{rson2}
\end{eqnarray}
For $\gamma=4/3$, $\rso$ is $0.25$ in units of the Bondi radius in the non-rotating limit. As discussed in Section~\ref{sec:method}, we adopt $\rso$ to be $1{,}500\,\mathrm{km}$, which is in line with the results of numerical simulations. This means that $r_\mathrm{B}=6000\,\mathrm{km}$ and $(GMr_\mathrm{B})^{1/2}$ is about $ 3\times10^{17}\,\mathrm{cm^2/s}$. Thus, for our adopted scale of $\rso$, the unit of $L$ is $ 3\times10^{17}\,\mathrm{cm^2/s}$. As discussed in Section~\ref{sec:method}, the highest angular momentum expected in the inner regions of massive stars is $3\times10^{16}\,\mathrm{cm^2/s}$, which corresponds to $L$ of $0.1$ in these units. We obtain the background solution by solving Eqs.~(\ref{eq:bondi1})-(\ref{eq:bondi2}) using numerical root finding. The speed of sound is then obtained using Eq.~(\ref{eq:c}), while the radial velocity is obtained using Eq.~(\ref{eq:v}).

\subsubsection{Derivatives of $v$, $c$, and ${\cal M}$}
\label{sec:derivatives}

As we will see below, the evolution of the perturbations depends on the derivatives of the background velocity $v$, sound speed $c$, and Mach number ${\cal M}$, which we calculate in this section. We differentiate Eq.~(\ref{eq:bondi2}) and divide the resulting equation by $r^2 \upsilon_r c^{\frac{2}{\gamma-1}}$, which leads to
\begin{equation}
2 + \frac{r}{\upsilon_r} \frac{\partial \upsilon_r}{\partial r} + \frac{2}{\gamma-1} \frac{r}{c }\frac{\partial c}{\partial r} = 0
\end{equation}
or
\begin{equation}
\label{eq:dlvdlc}
2 + \frac{\partial \log \upsilon_r}{\partial \log r} + \frac{2}{\gamma-1} \frac{\partial \log c}{\partial \log r} = 0
\end{equation}
The derivative of Eq.~(\ref{eq:bondi1}) is 
\begin{equation}
\upsilon_r \frac{\partial \upsilon_r}{\partial r} + \frac{2}{\gamma-1} c \frac{\partial c}{\partial r}= -\frac{1}{r^2} + \frac{L^2}{r^3}, 
\end{equation}
We rewrite it as 
\begin{equation}
\upsilon_r^2 \frac{\partial \log \upsilon_r}{\partial \log r}  + \frac{2}{\gamma-1} c^2 \frac{\partial \log c}{\partial \log r} = -\frac{1}{r} + \frac{L^2}{r^2}, 
\end{equation}
which we divide by $c^2$ and write as 
\begin{equation}
{\cal M}^2 \frac{\partial \log \upsilon_r}{\partial \log r}  + \frac{2}{\gamma-1} \frac{\partial \log c}{\partial \log r} = -\frac{1}{rc^2} + \frac{L^2}{r^2c^2}, 
\end{equation}
We can substitute equation (\ref{eq:dlvdlc}) into the last equation to obtain
\begin{equation}
{\cal M}^2 \frac{\partial \log \upsilon_r}{\partial \log r} - 2 - \frac{\partial \log \upsilon_r}{\partial \log r} = -\frac{1}{rc^2} + \frac{L^2}{r^2c^2}, 
\end{equation}
or
\begin{equation}
(1-{\cal M}^2) \frac{\partial \log \upsilon_r}{\partial \log r} + 2 = \frac{1}{rc^2} - \frac{L^2}{r^2c^2}, 
\end{equation}
from which we find 
\begin{equation}
\label{eq:dlvdlr}
\frac{\partial \log \upsilon_r}{\partial \log r} = \frac{1}{1-{\cal M}^2}\left( - 2 + \frac{1}{rc^2} - \frac{L^2}{r^2c^2} \right), 
\end{equation}
Finally, using equation (\ref{eq:dlvdlc}), we obtain
\begin{equation}
\label{eq:dlogv_dlogc}
\frac{\partial \log c}{\partial \log r} =  \frac{1-\gamma}{2}\left(
2 +\frac{\partial \log \upsilon_r}{\partial \log r} \right).
\end{equation}
For the Mach number at the sonic point, we obtain 
\begin{equation}
\label{eq:dlogM_dlogr}
\frac{\partial \log M}{\partial \log r} = \frac{\partial \log \upsilon_r}{\partial \log r} - \frac{\partial \log c}{\partial \log r}.
\end{equation}
For $L=0$, we recover the equations (A.3)-(A.4) of \cite{foglizzo:01}.

\section{Solution for the perturbations}
\label{sec:perturbation_solution}

The set of linearized equations after a Fourier transform in time is the following
\begin{flalign}
&-i\omega\delta\rho+\boldsymbol{\nabla}\cdot(\rho \delta \boldsymbol{\upsilon}+\boldsymbol{\upsilon} \delta\rho)=0,&\\
&-i\omega \delta \boldsymbol{\upsilon}+\delta \boldsymbol{w}\times \boldsymbol{\upsilon}+\boldsymbol{w}\times \delta \boldsymbol{\upsilon} 
&
\nonumber \\
&
\qquad
+\boldsymbol{\nabla}\left(\upsilon_r\delta \upsilon_r+\upsilon_\phi\delta \upsilon_\phi+{\delta c^2\over\gamma-1}\right)
={c^2\over\gamma} \boldsymbol{\nabla} \delta S,&\\
&-i\omega \delta S+\boldsymbol{\upsilon}\cdot\boldsymbol{\nabla} \delta S=0,&
\end{flalign}
where $\omega$ is time frequency.  
We define $\delta f$, $\delta h$ as follows:
\begin{eqnarray}
\delta f&\equiv&{\upsilon_r\delta \upsilon_r}+{L\over r}\delta \upsilon_\phi+ {\delta c^2\over\gamma-1},\\
\delta h&\equiv&{\delta \upsilon_r\over \upsilon_r}+{\delta \rho\over\rho}.
\end{eqnarray}
Using the definition of the entropy
\begin{eqnarray}
\label{eq:df_def2}
{\delta f\over c^2}-{L\over rc^2}\delta \upsilon_\phi&\equiv&\M^2{\delta \upsilon_r\over \upsilon_r}+{1\over\gamma-1}{\delta c^2\over c^2},\\
\label{eq:dh_def2}
\delta h+\delta S&\equiv&{\delta \upsilon_r\over \upsilon_r}+{1\over\gamma-1}{\delta c^2\over c^2}.
\end{eqnarray}
Other variables are deduced from $\delta f,\delta h, \delta S$, and $\delta \upsilon_\phi$:
\begin{eqnarray}
{\delta \upsilon_r\over \upsilon_r}&=&{1\over 1-\M^2}\left(\delta h+\delta S-{\delta f\over c^2} +{L\over rc^2}\delta \upsilon_\phi\right),\\
{\delta c^2\over c^2}&=&{\gamma-1\over 1-\M^2}\left\lbrack
{\delta f\over c^2} -{L\over rc^2}\delta \upsilon_\phi-
\M^2(\delta h+\delta S)\right\rbrack,\\
{\delta \rho\over\rho}&=&-{1\over 1-\M^2}\left(\M^2\delta h+\delta S-{\delta f\over c^2} 
+{L\over rc^2}\delta \upsilon_\phi\right),\\
{\delta p\over {\gamma p}}&=&-{1\over 1-\M^2} \nonumber \\
&\times&\left(\M^2(\delta h + \delta S) + (1-\M^2){\delta S \over \gamma} - {\delta f\over c^2} + {L\over rc^2}\delta \upsilon_\phi\right),\nonumber\\
\end{eqnarray}
where $p$ is the stationary pressure and $\delta p$ is the pressure perturbation. We assume the $\phi$-dependence of the form $\exp(im\phi)$, where $m$ is the angular wavenumber of the perturbation. 
From the definition of vorticity,
\begin{eqnarray}
\delta w_\theta&=&-{1\over r}{\p r\delta \upsilon_\phi\over\p r}+{im\over r}\delta \upsilon_r.
\end{eqnarray}
The time-dependence is separated by assuming $\exp(-i\omega t)$, where $\omega$ is the frequency. The Doppler shifted frequency is noted $\omega'$:
\begin{eqnarray}
\omega'\equiv \omega-{mL\over r^2}.
\end{eqnarray}
The perturbed linearized equation of continuity leads to
\begin{eqnarray}
-i\omega'\delta \rho +{1\over r^2}{\p\over \p r}{r^2\rho \upsilon_r\delta h}+{im\over r}{\rho \delta  \upsilon_\phi}=0.
\end{eqnarray}
The perturbed linearized Euler equation is written including the perturbation of the potential
\begin{eqnarray}
-i\omega\delta \upsilon_r +{L\over r}\delta w_\theta+{\p\delta f\over \p r}={c^2\over\gamma}{\p\delta S\over \p r},\\
-i\omega\delta \upsilon_\phi-\upsilon_r \delta w_\theta+{im\over r}\delta f={c^2\over\gamma}{im\over r}\delta S.\label{Euleraz2}
\end{eqnarray}
The perturbed linearized entropy equation reads
\begin{eqnarray}
\left({\p\over \p r}-{i\omega'\over \upsilon_r}\right)\delta S&=&0.\label{eqS2}
\end{eqnarray}
The vorticity equation follows from the curl of the Euler equation:
\begin{eqnarray}
{\p\over \p r}(r\upsilon_r\delta w_\theta)={i\omega' r}\delta w_\theta-im{\delta S\over \gamma}{\p c^2\over \p r}
.\label{eqvort2}
\end{eqnarray}
We define $\delta K$ as follows:
\begin{eqnarray}
\delta K&\equiv &-im r\upsilon_r\delta w_\theta + m^2 {c^2\over\gamma}\delta S.
\end{eqnarray}
This definition is the same as that in \cite{foglizzo:01} if we replace $m^2$ with $\ell(\ell+1)$. We note from Eqs.~(\ref{eqS2}) and (\ref{eqvort2}) that
\begin{eqnarray}
\left({\p\over \p r}-{i\omega'\over \upsilon_r}\right)\delta K&=&0.
\end{eqnarray}
The conservation of $\delta K$ in an adiabatic flow is thus also valid when self-gravity is included.
The differential system is expressed with the following five equations:
\begin{eqnarray}
{\p\delta f\over \p r}&=&i\omega\delta \upsilon_r +{c^2\over\gamma}{i\omega' \over \upsilon_r}\delta S-{L\over r}\delta w_\theta
,\\
{\p\delta h\over \p r}&=&{i\omega'\over \upsilon_r}{\delta \rho\over\rho} -{im\over r\upsilon_r}{\delta  \upsilon_\phi},\\
\left({\p\over \p r}-{i\omega'\over \upsilon_r}\right)\delta K&=&0,\\
\left({\p\over \p r}-{i\omega'\over \upsilon_r}\right)\delta S&=&0.
\end{eqnarray}
with, using Eq.~(\ref{Euleraz2})
\begin{eqnarray}
\label{eq:vtheta_f_dK1}
\omega r\delta \upsilon_\phi&=&-\frac{\delta K}{m} +{m}\delta f,\\
\label{eq:vtheta_f_dK2}
\delta f&=&{\omega\over m} r\delta \upsilon_\phi+{\delta K\over m^2}.
\end{eqnarray}
Thus $\delta S$ and $\delta K$ can be integrated as
\begin{eqnarray}
\delta S&=&\delta S_0\e^{\int_{r_0} {i\omega'\over \upsilon_r}\d r},\\
\delta K&=&\delta K_0\e^{\int_{r_0} {i\omega'\over \upsilon_r}\d r},\\
\delta w_\theta&= & {i\over r\upsilon_r}\left\lbrack \frac{\delta K_0}{m} - m{c^2\over\gamma}\delta S_0\right\rbrack \e^{\int_{r_0} {i\omega'\over \upsilon_r}\d r}. \label{eq:wtheta}
\end{eqnarray}
The variable $\delta f$ can be replaced by the variable $r\delta \upsilon_\phi$:
\begin{eqnarray}
{\p \over \p r}( r\delta \upsilon_\phi)&=&im\delta \upsilon_r - i {\delta K\over {m \upsilon_r}}+ im{c^2\over\gamma \upsilon_r}\delta S,\label{diff1}\\
{\p\delta h\over \p r}&=&{i\omega'\over \upsilon_r}{\delta \rho\over\rho} -{im\over r\upsilon_r}{\delta  \upsilon_\phi},\label{diff2}.
\end{eqnarray}
with
\begin{flalign}
&{\delta \upsilon_r\over \upsilon_r}={1\over 1-\M^2}\left(\delta h+\delta S-{\omega'\over mc^2} r\delta \upsilon_\phi-{\delta K\over m^2c^2} \right),& \label{dv/v}\\
&{\delta \rho\over\rho}=-{1\over 1-\M^2}\left(\M^2\delta h+\delta S-{\omega'\over mc^2} r\delta \upsilon_\phi-{\delta K\over m^2c^2}\right),& \label{drho/rho}\\
&{\delta p\over {\gamma p}}=-{1\over 1-\M^2} \times& \label{dp/p} \nonumber \\
&\left[ \M^2(\delta h+\delta S)+(1-\M^2){\delta S\over\gamma}-{\omega'\over mc^2}r\delta \upsilon_\phi-{\delta K\over m^2c^2}\right].&
\end{flalign}
We note that using the variable ($\delta \upsilon_\phi,\delta h,\delta S,\delta K$), the only appearance of the frequency $\omega$ and the angular momentum $L$ in the differential system (\ref{diff1})-(\ref{diff2}) is through the Doppler shifted frequency $\omega'$.
A single differential equation of second order can be obtained using the new variable $X$ in the adiabatic approximation. 
\begin{flalign}
\label{eq:du_system}
&\left({\p \over \p X}+{i\omega'\over c^2}\right)( r\delta \upsilon_\phi) \, =\, im \delta h & \nonumber \\ 
&\qquad \qquad
 -i{\delta K\over {m\upsilon_r^2}}+{im\over \gamma}\delta S\left({1\over \M^2}+\gamma-1\right),& \\
&
\label{eq:dh_system}
\left({\p \over \p X}+{i\omega'\over c^2}\right)\delta h \, = \, {i W\over m} r\delta \upsilon_\phi -{i\omega'\over \upsilon_r^2} \left(\delta S-{\delta K\over m^2c^2}\right).& 
\end{flalign}
where variable $X$ is related to $r$ via equation
\begin{eqnarray}
\label{eq:X}
{\p X\over \p r}\equiv {\upsilon_r\over 1-\M^2}.
\end{eqnarray}
The parameter $W$ is defined as
\begin{eqnarray}
W = \frac{\omega'^2\mu^2}{\upsilon_r^2 c^2},
\end{eqnarray}
while $\mu$ is defined as
\begin{eqnarray}
\mu^2\equiv 1-{m^2c^2\over r^2\omega'^2}(1-\M^2).
\end{eqnarray}
Thus
\begin{flalign}
&
\left\lbrace\left({\p \over \p X}+{i\omega'\over c^2}\right)^2+W \right\rbrace( r\delta \upsilon_\phi)
& \nonumber \\
& \qquad = {\omega'm\over \upsilon_r^2}
\left(\delta S-{\delta K\over m^2c^2}\right) & \nonumber \\
& \hspace{1.0cm}
-im\left({\p \over \p X}+{i\omega'\over c^2}\right)\left\lbrack
{\delta K\over m^2\upsilon_r^2}
-{\delta S\over \gamma}\left({1\over \M^2}+\gamma-1\right)\right\rbrack.&
\end{flalign}
Using Eq.~(\ref{eq:wtheta}), we can rewrite this equation as
\begin{flalign}
&
\left\lbrace\left({\p \over \p X}+{i\omega'\over c^2}\right)^2+W \right\rbrace( r\delta \upsilon_\phi)=
-{\p \over \p X}\left({r\delta w_\theta\over \upsilon_r}
\right)
, & \label{eq:diff_eq1}
\end{flalign}
It is remarkable that this equation is formally similar to the equation in an isentropic flow without rotation.
Again the effect of angular momentum is contained in $\omega'$. The source term responsible for the radial transport of angular momentum is fully described by the radial profile of perturbed vorticity rather than $\delta K$ or $\delta S$. Using variable
\begin{flalign}
\delta \tilde \upsilon_\phi\equiv \delta  \upsilon_\phi \e^{\int {i\omega'\over c^2}\d X}, \label{eq:vphitilde}
\end{flalign}
system (\ref{eq:du_system})-(\ref{eq:dh_system}) transforms into
\begin{flalign}
\label{eq:dut_system}
&{\p (r\delta \tilde \upsilon_\phi) \over \p X} = im \delta \tilde h & \nonumber \\ 
&\quad \qquad + \left[-i{\delta K\over {m\upsilon_r^2}}+{im\over \gamma}\delta S\left({1\over \M^2}+\gamma-1\right) \right] \e^{\int {i\omega'\over c^2}\d X} ,& \\
&
\label{eq:dht_system}
{\p \delta \tilde h \over \p X} = {i W\over m} r\delta \tilde \upsilon_\phi -{i\omega'\over \upsilon_r^2} \left(\delta S-{\delta K\over m^2c^2}\right) \e^{\int {i\omega'\over c^2}\d X},&  
\end{flalign}
while Eq.~(\ref{eq:diff_eq1}) becomes  
\begin{flalign}
\left\lbrace{\p^2 \over \p X^2}+W\right\rbrace( r\delta \tilde \upsilon_\phi)=-
\e^{\int {i\omega'\over c^2}\d X}{\p \over \p X}{r\delta w_\theta\over \upsilon_r}. \label{rdvtilde} 
\end{flalign}

\subsection{In- and out-going homogeneous solutions}
\label{sec:refraction_coeff}

We first re-write the homogeneous system in terms of $\delta \tilde f$ and $\delta \tilde h$: 
\begin{eqnarray}
\frac{\p \delta \tilde f}{\p X }
&=& i \omega \delta \tilde h \label{eq:dtux_hom} \\
\frac{\p \delta \tilde h}{\p X }
&=& \frac{i W}{\omega} \delta \tilde f \label{eq:dthx_hom} 
\end{eqnarray}
This system can solved analytically at large radii, where WKB approximation is valid. The solutions represent in-going and out-going acoustic waves \citep{foglizzo:01}: 
\begin{equation}
 \delta \tilde f^\pm = \tilde A_\pm \frac{\omega^{1/2}}{W^{1/4}} \exp\left(\pm i \int W^{1/2} dX \right). \label{eq:dupm_wkb}
\end{equation}
The solution that is regular at the sonic point is a linear combination of these solutions:
\begin{equation}
    \delta \tilde f_{0} = a \delta \tilde f^+ + b \delta \tilde f^-, \label{eq:du0}
\end{equation}
where $a$ and $b$ are some constants. Differentiation of equation (\ref{eq:du0}) leads to 
\begin{equation}
    \p_X \delta \tilde f_{0} = a \p_X \delta \tilde f^+ + b \p_X \delta \tilde f^-, \label{eq:dx_dtu0}
\end{equation}
Since $\p_X W^{1/2} \ll W$ in the WKB regime \citep{abdikamalov20}, the derivatives of functions $\delta \tilde f^\pm$ are
\begin{equation}
    \p_X \delta \tilde f^\pm = \pm i W^{1/2} \delta \tilde f^\pm. \label{eq:dxdtuphi}
\end{equation}
This equation allows us to obtain the Wronskien of the two solutions:
\begin{equation}
    {\cal W}_X = \p_X \delta \tilde f^+ \delta \tilde f^- - \p_X \delta \tilde f^+ = \tilde A_+ \tilde A_- {2 i \omega} 
\end{equation}
or, for variable $r$, 
\begin{equation}
    {\cal W}_r = \p_r \delta \tilde f^+ \delta \tilde f^- - \p_r \delta \tilde f^+ = \tilde A_+ \tilde A_- \frac{2 i \omega \upsilon_r}{1-M^2} 
\end{equation}
Using Eqs.~(\ref{eq:dtux_hom})-(\ref{eq:dthx_hom}) and (\ref{eq:dxdtuphi}), Eq.~(\ref{eq:dx_dtu0}) can be written as
\begin{equation}
     \frac{\omega}{W^{1/2}} \delta \tilde h_0 = a \delta \tilde f^+ - b \delta \tilde f^-.
\end{equation}
Combining this with Eq.~(\ref{eq:du0}), we obtain:
\begin{eqnarray}
\delta \tilde f_{0} + \frac{\omega}{W^{1/2}} \delta \tilde h_0 &=& 2 a \delta \tilde f^+, \\
\delta \tilde f_{0} - \frac{\omega}{W^{1/2}} \delta \tilde h_0 &=& 2 b \delta \tilde f^-,
\end{eqnarray}
from which we can easily obtain coefficients $a$ and $b$. The refraction coefficient is defined as ${\cal R}_s = b/a$ \citep{foglizzo:01}, which leads to
\begin{equation}
    {\cal R}_s = \frac{\delta \tilde f^+}{\delta \tilde f^-} \, \frac{\delta \tilde f_{0} - \frac{\omega}{W^{1/2}}\delta \tilde h_0}{\delta \tilde f_{0} + \frac{\omega}{W^{1/2}}\delta \tilde h_0}
\end{equation}

\subsection{Full inhomogeneous solution}
\label{sec:full_inho_sol}

In order to obtain the solution for advected vorticity waves ($\delta K \ne 0 $ and $\delta S =0$), we first re-write Eq.~(\ref{rdvtilde}) in terms of the function $\delta f$:
\begin{eqnarray}
\left\lbrace{\p^2 \over \p X^2}+W\right\rbrace \delta \tilde f=- \frac{1-{M}^2}{\upsilon_r} 
\e^{\int {i\omega'\over c^2}\d X}  J(r) \frac{\delta K}{r^2 \upsilon_r}, 
\label{eq:dftilde} 
\end{eqnarray}
where $J$ 
\begin{eqnarray}
J(r)=1-\frac{L}{m}\left(\frac{\omega'}{\upsilon_r^2}+i\frac{\p_r \upsilon_r^2}{\upsilon_r^3} +\frac{2i}{r\upsilon_r}\right)
\label{eq:j_of_r}
\end{eqnarray}
In the non-rotating limit, $J(r)=1$. Using the using the method of Green's functions, we can obtain the solution in the subsonic region \citep{abdikamalov20}
\begin{eqnarray}
\delta f(r>\rso)=-{i\delta K_R\over2\omega A_R} \times
\hspace{3.5cm} 
\nonumber \\ 
\bigg\{
\delta f^-\int_{\rso}^r\e^{i\alpha} A(r)\,\delta f_0\,\d r 
- 
\delta f_0\int_{\infty}^r\e^{i\alpha} A(r)\,\delta f^-\,\d r
\bigg\},
\label{eq:outsidem}
\end{eqnarray}
where
\begin{eqnarray}
\alpha = \int_R^r \omega' {1+\M^2\over 1-\M^2} \frac{\d r}{\upsilon_r}.   
\end{eqnarray}
and
\begin{eqnarray}
A(r) = {J(r)\over r^2 \upsilon_r}.   
\end{eqnarray}
The two free parameters of the solution are fixed by imposing the regularity at $r=\rso$ and assuming that no sound waves come from infinity as in \cite{abdikamalov20}. The second integral does not converge in rotating case since $J\gg1$ for $r\gg r_s$. In order to achieve convergence, we perform integration by parts of the integrals in (\ref{eq:outsidem}):
\begin{eqnarray}
&&{\delta f}(r>\rso)={\delta K_R\over2\omega A_R} \times \hspace{0cm} \nonumber \\ &&
\hspace{-1cm} \bigg\{
{\delta f^-}
\int_{r_{\rm s}}^{r}
\e^{i \alpha}
\left\lbrack
B(r) \delta f_0
+ C(r) \delta g_0
\right\rbrack\d r \nonumber\\
&& \hspace{-0.8cm} -{\delta f_0}
\int_{\infty}^{r}
\e^{i\alpha}
\left\lbrack \!
B(r) \delta f^- 
+ C(r) \delta g^-
\right\rbrack\d r
\bigg\},\nonumber \\ \label{outsidemnum}
\end{eqnarray}
where
\begin{eqnarray}
B(r) &=& {\p\over\p r}\left({1\!-\!\M^{2}\over r^{2}} \frac{J}{\omega'}\right),\\
C(r) &=& \frac{J}{\omega'} {i\omega \upsilon_r \over r^{2}}.
\end{eqnarray}
In the supersonic region, we obtain another homogeneous solution from a combination of $\delta f_0$ and the Wronskien (cf. Eq. C3 of \cite{abdikamalov20}). Once the homogeneous solutions are establishes, the full solution in the supersonic region, we again use the method of Green's function, as we did for obtaining the solution in the subsonic region (\ref{eq:outsidem}).

\subsection{Vortex velocities}
\label{sec:vorticity_solution}

We can analytically obtain the velocities of the vorticity waves without the contribution of the acoustic wave using the incompressibility condition, as described in Appendix F of \cite{abdikamalov20}. Combining equations (\ref{rdvtilde}) and (\ref{eq:wtheta}) for $\delta{S} = 0$ and $\delta{K} \ne 0$, we get a differential equation for $r\delta \tilde \upsilon_\phi$:
\begin{flalign}
\frac{\p^2}{\partial{X}^2}(r\delta\tilde{v}_\phi) + W(r\delta\tilde{v}_\phi) = A\frac{\delta{K_0}}{im}e^{\int \frac{i\omega'}{v^2}dX},
\end{flalign}
where $A \equiv {i\omega'(1-\M^2)}/{\upsilon_r^4}$. The solution of this equation is
\begin{flalign}
r\delta\tilde{v}_\phi = \frac{A}{W - \frac{\omega'^2}{v^4}}\frac{\delta{K_0}}{im}e^{\int \frac{i\omega'}{v^2}dX},
\end{flalign}
$r\delta{\upsilon_\phi}$ is obtained using transformation (\ref{eq:vphitilde}) 
\begin{flalign}
\label{eq:vphi_vor}
r\delta{\upsilon_\phi} = \frac{A}{W - \frac{\omega'^2}{v^4}}\frac{\delta{K}}{im} = - \frac{\delta{K}}{m\omega'}\frac{1-\M^2}{1-\M^2\mu^2}.
\end{flalign}
We use Eq.~(\ref{eq:vtheta_f_dK2}) to obtain
\begin{equation}
    \delta{f}^K = -\left[ \frac{\omega}{\omega'}\frac{1-M^2}{1-M^2\mu^2}-1 \right]\frac{\delta{K}}{m^2}.
\end{equation}
From Eq.~(\ref{eq:dh_system}), we obtain $\delta h$:
\begin{equation}
    \delta{h}^K = - \frac{1}{\upsilon_r^2} \frac{\M^2(\mu^2-1)}{1-\M^2\mu^2} \frac{\delta{K}}{m^2}.
\end{equation}
Finally, using Eq.~(\ref{dv/v}), we obtain 
\begin{equation}
\label{eq:vr_vor}
    \frac{\delta \upsilon_r}{\upsilon_r} = - \frac{1}{\upsilon_r^2} \frac{\M^2(\mu^2-1)}{1-\M^2\mu^2} \frac{\delta{K}}{m^2}.
\end{equation}

\end{document}